\documentclass[aps,showpacs,preprintnumbers,referee, superscriptaddress, twocolumn,floatfix,referee]{revtex4-2}
\usepackage{amsmath}
\usepackage{eurosym}
\usepackage{amssymb}
\usepackage{graphicx}
\usepackage{color}

\newcommand{\bea}{\begin{eqnarray}}
\newcommand{\eea}{\end{eqnarray}}
\newcommand{\be}{\begin{equation}}
\newcommand{\ee}{\end{equation}}
\newcommand{\p}{\prime}

\newcommand{\f}[2]{\frac{#1}{#2}}

\begin{document}

\title{Generalizing the coupling between geometry and matter: $f\left(R,L_m,T\right)$ gravity}
\author{Zahra Haghani}
\email{z.haghani@du.ac.ir}
\affiliation{ School of Physics, Damghan University, Damghan, Iran}
\author{Tiberiu Harko}
\email{tiberiu.harko@aira.astro.ro}
\affiliation{Astronomical Observatory, 19 Ciresilor Street,  Cluj-Napoca 400487, Romania, }
\affiliation{Department of Physics, Babes-Bolyai University, Kogalniceanu Street,
Cluj-Napoca 400084, Romania,}
\affiliation{School of Physics, Sun
Yat-Sen University, Guangzhou 510275, People's Republic of China}

\begin{abstract}
We generalize and unify the $f\left(R,T\right)$ and $f\left(R,L_m\right)$
type gravity models by assuming that the gravitational Lagrangian is given
by an arbitrary function of the Ricci scalar $R$, of the trace  of the
energy-momentum tensor $T$, and of the matter Lagrangian $L_m$, so that $%
L_{grav}=f\left(R,L_m,T\right)$. We obtain the gravitational field equations
in the metric formalism,  the equations of motion for test
particles, and the energy and momentum balance equations, which follow from the covariant divergence of the energy-momentum
tensor. Generally, the motion is non-geodesic, and takes place in the
presence of an extra force orthogonal to the four-velocity. The Newtonian limit of the equations of motion is also investigated, and the expression of the extra acceleration is obtained for small velocities and weak gravitational fields. The generalized Poisson equation is also obtained in the Newtonian limit, and the Dolgov-Kawasaki instability is also investigated.   The cosmological implications of the theory are investigated for a homogeneous, isotropic and flat Universe for two particular choices of the Lagrangian density $f\left(R,L_m,T\right)$ of the gravitational field, with a multiplicative and additive algebraic structure in the matter couplings, respectively, and for two choices of the matter Lagrangian, by using both analytical and numerical methods.
\end{abstract}

\pacs{04.50.Kd,04.20.Cv, 95.35.+d}
\date{\today}
\maketitle
\tableofcontents

\section{Introduction}

The unexpected discovery of the accelerating behavior of the Universe \cite%
{1n,2n,3n,4n,acc}, that began at a redshift $z$ of around $z\approx 0.6$,  brought into question the very foundations of the most successful gravitational theory developed so far, general relativity. The simplest answer to explain the late time acceleration is to resort again to the old cosmological constant $\Lambda$, which was introduced by Einstein to build the first general relativistic cosmological model \cite{Ein}.  Based on this ansatz, and strongly supported by the excellent fits a cosmological constant provide to the observations, the standard model of cosmology, called the $\Lambda$CDM paradigm, was established, which also requires the existence of another intriguing (and yet undetected) constituent of the Universe, called dark matter \cite{Sal}. The $\Lambda$CDM model describes well observations \cite{C1,C2,C3, C4}, however, it is confronted with a major problem:  no fundamental physical
theory can give a reason for it. The main problem arises from the difficulties encountered in explaining the nature and origin of the cosmological constant \cite{Wein1,Wein2, Wein3}. A popular explanation for $\Lambda$ as a Planck-scale vacuum energy density $\rho _{vac}$ leads to the ``worst prediction in physics" \cite{Lake}, since $\rho _{vac}\approx \left(\hbar/c\right)\int _{k_{dS}}^{k_{Pl}}{\sqrt{k^2+\left(mc/\hbar\right)^2}d^3k}\approx \rho _{Pl}=c^5/\hbar G^2=10^{93}\; {\rm g/cm^3}$, a predictions that differs by a factor of around $10^{-120}$  from the observed value of the cosmological constant, $\rho _{\Lambda}=\Lambda c^2/8\pi G\approx 10^{-30}\;{\rm g/cm^3}$ \cite{C3}.

Actually, the $\Lambda $CDM model can fit the
cosmological observational data at a very high level of precision. Additionally, it is a very
simple (empirical in some sense) general relativistic theoretical approach to cosmology. However, even at the observational level it faces some important problems, the most important being the ``Hubble tension", representing the significant differences between the values of the Hubble constant, $H_0$, as
determined from the CMB measurement \cite{C4}, and those obtained from  direct measurements in the local Universe \cite{M1,M2,M3}. For example
the SH0ES measurement of $H_0$ give
$H_0 = 74.03 \pm 1.42$ km/s/Mpc \cite{M1},  while from the early Universe
measurement of the Planck satellite one finds $H_0 = 67.4 \pm 0.5$ km/s/Mpc \cite{C3}, results
which differ by $\sim  5\sigma$.

On the other hand, important theoretical questions cannot be also answered in the
framework of the $\Lambda$CDM paradigm. Why it happens that the numerical value of the cosmological constant is extremely small? Why is its value adjusted so fine? What determined the Universe to begin to accelerate only recently, after a long decelerating phase? And, after all, is a cosmological constant really necessary for theoretical and observational cosmology?
Hence, the present day observational situation of cosmology strongly requires the exploration of alternative pathways for the understanding of the gravitational interaction, and its influence on the dynamics of the Universe. In fact, one can distinguish (at least) three possible theoretical approaches  as substitutes of the $\Lambda$CDM paradigm.

The first avenue leading to an alternative description of cosmological evolution may be called the {\it dark constituents model}. It is based on the generalization of the Einstein gravitational field equations through the addition of two new terms in the total energy momentum tensor of the Universe, which describe dark energy and dark matter, respectively. In this approach gravitational phenomena are described by the generalized field equation \cite{HL20}
\be
G_{\mu \nu}=\kappa ^2 T^{\rm bar}_{\mu \nu}+\kappa ^2T^{\rm DM}_{\mu \nu}(\phi, \psi _{\mu},...)+\kappa ^2T^{\rm DE}_{\mu \nu}(\phi, \psi _{\mu},...),
\ee
where $G_{\mu \nu}$ is the Einstein tensor, $T^{\rm bar}_{\mu \nu}$, $T^{\rm DM}_{\mu \nu}(\phi, \psi_{\mu},...)$, and $T^{\rm DE}_{\mu \nu}(\phi, \psi_{\mu},...)$ denote the energy-momentum tensors of baryonic matter, dark matter and dark energy, respectively.  In this approach dark energy is understood as a form of matter, an interpretation based on the equivalence between mass and energy in the theory of relativity.  Both the energy-momentum tensors of dark matter and dark energy can be constructed from some scalar or vector fields.  The simplest dark constituent model can be obtained by assuming that dark energy is a scalar field $\phi$ with a self-interaction potential $V(\phi)$. Hence the action of the gravitational theory becomes $S=\int{\left[M_{Pl}^2R/2-\left(\partial \phi\right)^2-V(\phi)\right]\sqrt{-g}d^4x}$, leading to so-called quintessence models \cite{Q1,Q2,Q3,Q4,Q5, Q6}.  Other dark constituent models are k-essence models \cite{K1,K2,K3}, tachyon field models \cite{T1,T2}, models considering phantom fields \cite{Ph1,Ph2,Ph3}, or quintom field models \cite{Qu1,Qu2,Qu3}, respectively. In \cite{Q6} it was pointed out  that quintessence always lower $H_0$ relative to the $\Lambda$CDM model, thus making the Hubble tension worse. 
Chameleon field \cite{Ch1,Ch2,Ch3,nonlocal,Ch4}, Chaplygin gas \cite{Cha1,Cha2} and vector field \cite{V1,V2,V3} dark energy  models
have also been investigated.  Notwithstanding its impressive success, the dark constituents avenue is still confronted with its own intrinsic problems, the most important being the lack of any direct experimental evidence for the existence of dark matter.  Furthermore, the existence of so many distinct field type models for dark energy, each one facing its own theoretical problems, leads to the problem of the existence of a single,  theoretically consistent field theoretical picture of the major components of the Universe. For reviews of the dark energy models see \cite{Rev1,Rev2,Rev3,Rev4}.

The second pathway to gravitational phenomena is the so-called {\it dark gravity} approach. The dark gravity avenue, going beyond the Einsteinian perspective, is based on a purely geometrical description of the gravitational phenomena, with their dynamics and evolution explained as a result of the  modification of the geometric format of the Einstein field equations. In the dark gravity approach, the Einstein equations are generally formulated as
\be
G_{\mu \nu}=\kappa ^2T_{\mu \nu}^{(mat)}+\kappa ^2 T_{\mu \nu}^{(\rm geom)}\left(g_{\mu \nu}, R, \square R,...\right),
\ee
 where $T_{\mu \nu}^{(mat)}$ is the matter energy-momentum tensor, while $T_{\mu \nu}^{(\rm geom)}\left(g_{\mu \nu}, R, \square R,...\right)$, playing the role of an effective energy-momentum tensor, is a geometric quantity, constructed entirely from the metric.  $T_{\mu \nu}^{(\rm geom)}\left(g_{\mu \nu}, R, \square R,...\right)$  may mimic dark energy, dark matter, or both. An important example of dark gravity is the $f(R)$ theory, introduced initially in \cite{Bu1}, and developed in its early stages in \cite{Bu2,Bu3,Bu4,Bu5}. In $f(R)$ gravity theory the standard Hilbert-Einstein action $S=\int{\left(R/\kappa ^2+L_m\right)\sqrt{-g}d^4x}$ is replaced by a more general action given by $S= \int{\left[f(R)/\kappa^2+L_m\right]\sqrt{-g}d^4x}$, where $f(R)$ is an arbitrary analytical function of the Ricci scalar $R$. The astrophysical and cosmological implications of the $f(R)$ gravity have been extensively investigated \cite{fR1,fRn1,fRn2,fRn3,fR2,fR3,fR4,fR5,fR6,fRn4, fR7,fR8,fR9,fR10,fR11}. The first internally consistent $f(R)$ gravity model, describing both inflation, as well as present day dark energy, was constructed in \cite{fRn4}. Another geometric theory,  the hybrid metric-Palatini gravity theory \cite{HMP1,HMP2,HMP3,HMP4} extends and unifies two geometric formalisms, the metric and the Palatini ones, respectively. Theories involving geometric structures that go beyond the Riemannian one, like, for example, Weyl geometry, have also been investigated \cite{W1,W2,W3,W4,W5,W6,W7,W8,W9,W10,weylcartan}. For reviews of modified gravity theories, and their applications, see \cite{R1,R2,R3,R4, R5}.

A third way for understanding gravitation has also been developed, which we may call {\it the dark coupling approach}.  The basic idea of this approach is the substitution of the standard Hilbert-Einstein Lagrangian density, which has a simple additive structure in curvature and matter Lagrangian,  with a more general algebraic system. In this way {\it the maximal extension of the standard Hilbert-Einstein gravitational Lagrangian} can be established by considering that the gravitational action is an arbitrary analytical function of the curvature scalar $R$, of the matter Lagrangian $L_m$, and, possibly, of other thermodynamic parameters. This approach leads naturally to the existence of a {\it nonminimal coupling between geometry and matter}. Many types of coupling between matter and geometry are possible, including the coupling between the Ricci scalar and the trace of the matter energy-momentum tensor.

Therefore, in the dark coupling approach,  the Einstein gravitational equations can be extended to the form
\bea
G_{\mu \nu}=\kappa ^2T_{\mu \nu}^{(mat)}+
\kappa ^2 T_{\mu \nu}^{(\rm coup)}\left(g_{\mu \nu},  R, L_m, T, \square R, \square T,... \right),\nonumber\\
\eea
 with the effective energy-momentum tensor of the theory $T_{\mu \nu}^{(\rm coup)}\left(g_{\mu \nu}, R, L_m, T, \square R, \square T,... \right)$ is obtained by considering generally the maximal extension of the Hilbert-Einstein Lagrangian, and by introducing a {\it non-additive geometry-matter algebraic structure} that describes the couplings between matter in all its forms, and space-time curvature, respectively. It is important to mention that in both dark gravity and dark coupling theories {\it the gravitational constant $\kappa ^2$ turns into an effective quantity}, generally a function of the couplings, and of the field parameters.

 The dark coupling type theories were initiated in \cite{fLm1}, where a gravitational action of the form $S=\int{\left[f_1(R)+\left(1+\lambda f_2(R)\right)L_m\right]\sqrt{-g}d^4x}$ was investigated. This action was extended in \cite{fLm2}, finally leading to the $f\left(R,L_m\right)$ gravity theory \cite{fLm3}, with the gravitational action given by an arbitrary function of the Ricci scalar and of the matter Lagrangian, $S=\int{f\left(R,L_m\right)\sqrt{-g}d^4x}$. Different astrophysical and cosmological applications of the $f\left(R,L_m\right)$, as well as fundamental aspects of the theory were considered in \cite{fLm4, fLm5,fLm6,fLm6a, fLm6b, fLm7,fLm8,fLm9,fLm10,fLm11, fLm12}.

 An alternative possibility of coupling between matter and geometry is described by the $f(R,T)$ gravity theory \cite{fT1}, in which the trace of the matter energy-momentum tensor is nonminimally coupled to geometry, with the action given by $S=\int{\left[f\left(R,T\right)+L_m\right]\sqrt{-g}d^4x}$. The astrophysical and cosmological implications of $f(R,T)$ theory were investigated in detail in \cite{fT2,fT3,fT4,fT5,fT6,fT7,fT8,fT9,fT10,fT11,fT12,fT12a, fT13,fT13a, fT14, fT14a, fT14b, fT15, fT15a, fT16, fT17, fT18, fT19, fT20, fT21, fT22}.

The viability of the $f(R, T)$ cosmological models were discussed in \cite{fT12a} for a Lagrangian density of the form $f(R, T) = R + \lambda T + \gamma _nT^n$, where $\gamma _n$ and $n$ are constants. Several models were considered, corresponding to different values of the model parameters, and confronted with the $H(z)$ data and  Supernovae data. The $H(z)$ comparison indicates consistency with $\Lambda$CDM at low redshifts, but a significant deviation of the models from the standard scenario does appear when going to higher redshifts. Similar pathological behaviors also show up in the dynamics of the deceleration parameter of the models, leading to universes where
a transition from a decelerating phase to an accelerated one does not exist. The existence of such a transition is a basic requirement for any cosmological model compatible with the observational data. However, one can find models with transitions from deceleration to acceleration transition, but the predictions of these models are not in agreement with
both the $H(z)$ and Supernovae data at higher redshift regimes. On the other hand, at lower redshifts there is a good concordance between model and data.
These results indicate the difficulties faced by the considered version of the $f(R, T)$ theory in
 correctly representing the cosmological evolution in its entirety.

 A detailed analysis of the cosmological model based on the field equations $R_{\mu\nu}-(1/2)Rg_{\mu \nu}=[(8\pi +\lambda)/\lambda]T_{\mu \nu}+(p+T/2)g_{\mu \nu}$ was performed in \cite{fT14a}. The maximum likelihood analysis was used to obtain the constraints on the Hubble parameter $H_0$ and the model parameter by taking into account the observational Hubble data set $H(z)$, the Union 2.1 compilation data set SNeIa, the Baryon Acoustic Oscillation data BAO, and the joint data set $H(z)$ + SNeIa and $H(z)$ + SNeIa + BAO, respectively. It was found that the model is in good agreement with these observations.

 An extension  of the $f(R,T)$ gravity theory by introducing higher derivatives matter fields was proposed in \cite{fT23}, based on an action of the form $S=(1/16\pi)\int{f(R,T,\Box T)\sqrt{-g}d^4x}+\epsilon \int{L_m\sqrt{-g}d^4x}$, where $\epsilon =\pm 1$. Accelerated evolution does appear in this scenario in the dust-filled Universe without any additional matter sources. The inflationary model is also in quite good agreement with observations.

Gravitational theories in which the Ricci scalar is coupled to the square of the energy-momentum tensor $T^2=T_{\mu \nu}T^{\mu \nu}$, with action of the form $S=\int{f\left(R,T^2\right)\sqrt{-g}d^4x}$ were analyzed in \cite{B1,B2,B3,B4, B5}.  Theoretical models involving couplings between the Ricci tensor $R_{\mu \nu}$ and the energy momentum tensor $T^{\mu \nu}$, extending the action of the standard $f(R,T)$ theory to an action of the form $S=\left(1/8\pi G\right)\int{\left[f\left(R,T, R_{\mu \nu}T^{\mu \nu}\right)+L_m\right]\sqrt{-g}d^4x}$ were introduced and discussed in \cite{G1,G2,G3}. Derivative matter couplings are also considered in \cite{derivativematter}, where the Galileon interactions are promoted to contain matter Lagrangians. Nonminimal couplings between nonmetricity and matter were investigated in \cite{Qc1,Qc2,Qc3,Qc4}. For reviews and detailed discussions on the gravitational theories with geometry-matter coupling see \cite{book} and \cite{review}, respectively.

 It is the goal of the present paper {\it to extend, generalize, and at the same time unify two classes of gravitational theories with geometry-matter coupling, namely, the $f\left(R,L_m\right)$ and the $f(R,T)$ theories}, respectively. Hence we will consider a gravitational theory in which matter is non-minimally coupled with geometry at both the levels of the matter Lagrangian $L_m$, and of the trace of the energy-momentum tensor $T$, with Lagrangian density given by $f\left(R,L_m,T\right)$. This theory has as limiting cases the $f(R)$, $f\left(R,L_m\right)$ and $f(R,T)$ gravity theories, respectively.

 After introducing the gravitational action, as a first step in our study, we obtain the gravitational field equations of the theory. The energy and momentum balance equations are also obtained, and, as usual in theories with geometry-matter coupling, it turns out the matter energy-momentum tensor is not conserved anymore. The Newtonian limit of the equations of motion is investigated in detail, and the expression of the extra-acceleration is obtained for small particle velocities and weak gravitational fields. The Dolgov-Kawasaki instability is investigated in detail, and the condition for the existence of stable perturbations is obtained. The cosmological applications of the theory are also investigated. We consider two particular choices of the Lagrangian density $f\left(R,L_m,T\right)$ of the field, having a multiplicative and additive algebraic structure in the matter couplings. We also investigate the effects of the two possible different choices of the matter Lagrangian ($L_m=p$ and $L_m=-\rho$, respectively) on the description of the cosmological evolution.

 The present paper is organized as follows. The gravitational field equations of the $f\left(R,L_m,T\right)$ theory are obtained in Section~\ref{2}. The divergence of the matter energy-momentum tensor is derived in Section~\ref{3}. The energy and momentum balance equations are also presented, and it is shown that the equation of motion of test particles is not geodetic, but it takes place in the presence of an extra-force. The Newtonian limit of the equation of motion is investigated by using a variational approach, and the general expression of the extra-acceleration is obtained in the limit of small velocities. The generalized Poisson equation for the Newtonian gravitational potential is also derived.   The Dolgov-Kavasaki instability of the theory is analyzed in Section~\ref{4}, and the condition for obtaining stable perturbations is found. The cosmological implications of the theory are examined in Section~\ref{5}, for two specific choices of the function $f\left(R,L_m,T\right)$, which assumes a multiplicative and an additive form in the matter couplings. The cosmological evolution equations are studied for the two forms of the matter Lagrangian, $L_m=-\rho$, and $L_m=p$, respectively. The possibility of the existence of a de Sitter type phase for the models is explored.  We discuss and conclude our results in Section~\ref{7}.

\section{Field equations of $f\left(R,L_m,T\right)$ gravity}\label{2}

We assume that the action describing the modified gravity theory with strong
geometry-matter coupling takes the form
\begin{equation}
S=\frac{1}{16\pi }\int f\left( R,L_{m},T\right) \sqrt{-g}\;d^{4}x+\int {L_{m}%
\sqrt{-g}\;d^{4}x}\,
\end{equation}%
where $f\left( R,L_{m},T\right) $ is an arbitrary function of the Ricci scalar
$R$, of the trace $T$ of the matter energy-momentum tensor $T_{\mu \nu }$, $%
T=g^{\mu \nu }T_{\mu \nu }$, and of the Lagrangian density of the ordinary
matter, $L_{m}$, respectively. The main physical motivation for considering this function
is that it will ensure that the modification is non-trivial for all kinds of
matter fields, including radiation with $T=0$.

The energy-momentum tensor of the ordinary (baryonic) matter is defined in the standard way  according to the
relation \cite{LaLi}
\begin{equation}
T_{\mu \nu }=-\frac{2}{\sqrt{-g}}\frac{\delta \left( \sqrt{-g}L_{m}\right) }{%
\delta g^{\mu \nu }}\,.
\end{equation}

In the following we introduce the important assumption that the Lagrangian density $L_{m}$ of the matter
depends only on the metric tensor components $g_{\mu \nu }$, and not
on its derivatives. Therefore for $T_{\mu \nu}$ we find the simple expression
\begin{equation}  \label{en1}
T_{\mu \nu }=g_{\mu \nu }L_{m}-2\frac{\partial L_{m}}{\partial g^{\mu \nu }}.
\end{equation}

The variation of the action $S$ with respect to the metric tensor components
$g^{\mu \nu }$ can be obtained as
\begin{align}  \label{eq1}
\delta S&=\frac{1}{16\pi }\int \bigg[ f_{R}\delta R+\bigg(f_{T} \frac{\delta
T}{\delta g^{\mu \nu }}+f_{L} \frac{\delta L_{m}}{\delta g^{\mu \nu }}
\notag \\
&-\frac{1}{2}g_{\mu \nu }f -8\pi T_{\mu\nu}\bigg)\delta g^{\mu \nu }\bigg]
\sqrt{-g}\,d^{4}x\,,
\end{align}
where we have denoted $f_{R} =\partial f/\partial R$, $f_{T} =\partial f
/\partial T$, and $f_{L} =\partial f /\partial L_{m}$, respectively. The
variation of the Ricci scalar can be obtained from the Palatini identity $ \delta R_{\sigma\nu} = \nabla_\rho (\delta \Gamma^\rho_{\nu\sigma}) - \nabla_\nu (\delta \Gamma^\rho_{\rho\sigma})$ \cite{Ts} as
\begin{equation}
\delta R=R_{\mu \nu }\delta g^{\mu \nu }+g_{\mu \nu }\nabla _{\rho }\nabla
^{\rho }\delta g^{\mu \nu }-\nabla _{\mu }\nabla _{\nu }\delta g^{\mu \nu }.
\end{equation}
where $\nabla _{\lambda }$ represents the covariant derivative with respect
to the connection $\Gamma $, associated to the metric $g$, while $\Gamma
_{\mu \nu }^{\lambda }$ denotes the Christoffel symbols associated to the metric. We also assume the metric compatibility of the covariant derivative, $\nabla_\sigma g^{\mu\nu} = 0$.

The variation of the trace $T=T_{\mu}^{\mu}$ of the matter energy-momentum tensor $T_{\mu \nu}$ with
respect to the metric tensor is defined as
\begin{equation}
\frac{\delta \left( g^{\alpha \beta }T_{\alpha \beta }\right) }{\delta
g^{\mu \nu }}=T_{\mu \nu }+\Theta _{\mu \nu }\,,
\end{equation}%
where
\begin{equation}
\Theta _{\mu \nu }\equiv g^{\alpha \beta }\frac{\delta T_{\alpha \beta }}{%
\delta g^{\mu \nu }}=L_{m}g_{\mu \nu }-2T_{\mu \nu }-\tau _{\mu \nu },
\end{equation}%
and we have denoted
\begin{equation}
\tau _{\mu \nu }=2g^{\alpha \beta }\frac{\partial ^{2}L_{m}}{\partial g^{\mu \nu
}\partial g^{\alpha \beta }}.
\end{equation}

After substituting the above expressions into Eq.~\eqref{eq1}, and integrating by
parts, one can obtain the field equation of the $f(R,T,L_{m})$ gravity as
\begin{align}\label{eq2}
& f_{R}R_{\mu \nu }-\frac{1}{2}\left[ f-(f_{L}+2f_{T})L_{m}\right] g_{\mu
\nu }  \notag \\
& +\left( g_{\mu \nu }\Box -\nabla _{\mu }\nabla _{\nu }\right) f_{R}=\left[
8\pi +\frac{1}{2}(f_{L}+2f_{T})\right] T_{\mu \nu }  \notag \\
& +f_{T}\tau_{\mu\nu}.
\end{align}%
%
%
%
%

For perfect fluid matter Lagrangian with $L_{m}=-\rho $, and also for the
scalar field theory with $L_{m}=-\frac{1}{2}\partial _{\mu }\phi \partial
^{\mu }\phi +V(\phi )$, one can immediately see that $\tau_{\mu\nu}$ vanishes.
However, in the case of vector field theory, with Lagrangian $L_{m}=-\frac{1}{%
4}F_{\mu \nu }F^{\mu \nu }+f(A^{2})$, we obtain
\begin{equation}
\tau _{\mu \nu }=F_{\alpha\mu}F_{\nu
}^{~\alpha }+2A_{\mu }A_{\nu }A^{2}f^{\prime\prime}(A^{2}).
\end{equation}%

If $f=R$ (the Hilbert-Einstein Lagrangian), from Eq.~(\ref{eq2}) we recover
the standard field equations of general relativity, $R_{\mu \nu }-(1/2)g_{\mu \nu
}R=8\pi T_{\mu \nu }$ \cite{LaLi}. If $f=f\left(R,L_m\right)$, we reobtain the field equations of the $%
f\left(R,L_m\right)$ gravity model \cite{fLm1}, while $f=f\left( R,T\right) $ gives the field
equations of the $f\left( R,T\right) $ theory \cite{fT1}.

By taking the trace of the gravitational field equations (\ref{eq2}), in the
absence of electromagnetic type fields we obtain
\begin{align}\label{tr}
 f_{R}R&-2\big[ f-(f_{L}+2f_{T})L_{m}\big] +3\Box f_{R}  \nonumber \\
& =\left[ 8\pi +\frac{1}{2}\left( f_{L}+2f_{T}\right) \right] T+f_T\tau .
\end{align}

With the use of Eq. (\ref{tr}) the gravitational field equation of the $%
f\left( R,T,L_{m}\right) $ gravity theory can be written in an alternative
form as
\begin{eqnarray}\label{13}
R_{\mu \nu }-\frac{1}{4}Rg_{\mu \nu }&=&\frac{1}{f_{R}}\left[ 8\pi +\frac{1}{%
2}(f_{L}+2f_{T})\right] \left( T_{\mu \nu }-\frac{1}{4}g_{\mu \nu }T\right)
\notag \\
&&- \frac{1}{f_{R}}\left( \frac{1}{4}g_{\mu \nu }\Box -\nabla _{\mu }\nabla
_{\nu }\right) f_{R}  \notag \\
&&+\frac{f_T}{f_R}\left(\tau _{\mu \nu}-\frac{1}{4}\tau g_{\mu \nu}\right).
\end{eqnarray}

We call this form of the gravitational field equations of the $%
f\left(R,T,L_m\right)$ the \textit{traceless representation} of the theory. It is interesting to note that for $f(R)=R$, Eq.~(\ref{13}), the basic field equation of the traceless representation of the $f\left(R,L_m,T\right))$ reduces to the {\it geometry-matter symmetric Einstein equations} \cite{Ein1,Pauli},
\be\label{eqsym}
R_{\mu \nu }-\frac{1}{4}Rg_{\mu \nu }=8\pi G \left(T_{\mu \nu }-\frac{1}{4}g_{\mu \nu }T\right),
\ee
which have been proposed as a possible solution to the cosmological constant problem. For the applications and further generalizations of the geometry-matter symmetric Einstein equations see \cite{V3}.

\section{Energy and momentum balance equations, equations of motion, and
extra-force in $f\left(R,L_m,T\right)$ gravity}\label{3}

It is obvious that due of the non-minimal coupling between matter and
geometry, in the present theory the energy-momentum tensor is not conserved.
In this Section we will first obtain the non-conservation equation of the
matter energy-momentum tensor (the energy and momentum balance equations), and then we will consider the equations of
motion of particles, thus obtaining the expression of the extra-force generated by the curvature-matter
coupling.

\subsection{The divergence of the matter energy-momentum tensor}

Let us start with taking the covariant derivation of the gravitational field equations Eq. %
\eqref{eq2}
\begin{eqnarray}
&&R_{\mu \nu }\nabla ^{\mu }f_{R}+f_{R}\nabla ^{\mu }R_{\mu \nu }-\frac{1}{2}%
\nabla _{\nu }f=  \notag \\
&&\left[ 8\pi +\frac{1}{2}(f_{L}+2f_{T})\right] \nabla ^{\mu }T_{\mu \nu }
\notag \\
&&+(\Box \nabla _{\nu }-\nabla _{\nu }\Box )f_{R}-\nabla _{\nu }\Bigg[%
L_{m}\left( f_{T}+\frac{1}{2}f_{L}\right) \Bigg]  \notag \\
&&+T_{\mu \nu }\nabla ^{\mu }\left( f_{T}+\frac{1}{2}f_{L}\right) +A_{\nu },
\end{eqnarray}%
where we have defined
\begin{equation}
A_{\nu }\equiv \nabla ^{\mu }\left( f_{T}\tau_{\mu\nu}\right) .
\end{equation}%

One should note that in the case of perfect fluid and scalar field theory we
have $A_{\nu }=0$. By taking into account that
\begin{equation}
\nabla _{\nu }f\left( R,L_{m},T\right) =f_{R}\nabla _{\nu }R+f_{T}\nabla
_{\nu }T+f_{L}\nabla _{\nu }L_{m},
\end{equation}%
and after using the geometric identities $\nabla ^{\mu }G_{\mu \nu }=0$, and \cite{Ko06}
\begin{equation}
\left( \Box \nabla _{\nu }-\nabla _{\nu }\Box \right) f_{R}=R_{\mu \nu
}\nabla ^{\mu }f_{R},
\end{equation}%
the non-conservation equation of the energy-momentum tensor is reduced to
\begin{align}\label{noncons}
\nabla ^{\mu }T_{\mu \nu } &=\frac{1}{8\pi +f_{m}}\Big[\nabla _{\nu }\big(%
L_{m}f_{m}\big)-T_{\mu \nu }\nabla ^{\mu }f_{m}-A_{\nu
}\nonumber\\
&-\f12(f_T\nabla_\nu T+f_L\nabla_\nu L_m)\Big],
\end{align}%
where we have defined
\begin{equation}
f_{m}=f_{T}+\frac{1}{2}f_{L}.
\end{equation}%
Eq.~(\ref{noncons})  is direct a consequence of the presence of the matter fields in the expression of the gravitational Lagrange density, given by the function $f\left(R,L_m,T\right)$. One can immediately
see that in the case of $f_{T}=0=f_{L}$, the matter content of the Universe
is conserved.

By using the traceless representation of the field equations, one can obtain the non-conservation of the energy-momentum tensor as
\begin{align}\label{conssim}
\nabla ^{\mu }T_{\mu \nu } &=\frac{1}{4}\nabla _{\nu }T-\frac{1}{\left(
	8\pi +f_{m}\right) }\nabla
^{\mu }(f_TS_{\mu \nu })\nonumber\\&-\left( T_{\mu \nu }-\frac{1}{4}g_{\mu \nu }T\right) \nabla ^{\mu }\ln
\left( 8\pi +f_{m}\right)\notag \\
&+\frac{1}{4\left(
	8\pi +f_{m}\right) }\left(Rf_{R}\nabla _{\nu }\ln \frac{R}{f_{R}}-3\nabla _{\nu }\square f_{R}\right),
\end{align}%
where we have denoted
\begin{equation}
S_{\mu \nu }= \tau _{\mu \nu }-\frac{1}{4}\tau g_{\mu \nu }.
\end{equation}

\subsection{The energy and momentum balance equations}

In the following we adopt the simplifying assumption according to which the matter content of the Universe consists
of a perfect fluid that can be described by only two basic thermodynamics parameters, the energy density $\rho $, and the thermodynamic pressure $p$ of the fluid, respectively. In this case the matter energy-momentum tensor is given by
\begin{equation}
T_{\mu \nu }=\left( \rho +p\right) u_{\mu }u_{\nu }+pg_{\mu \nu },
\label{em}
\end{equation}%
where the four-velocity $u_{\mu }$ is normalized as $u^{\mu }u_{\mu }=-1$.
Next, we introduce the projection operator $h_{\lambda }^{\nu }$, defined as
\be
h_{\lambda }^{\nu }=\delta _{\lambda }^{\nu }+u_{\lambda }u^{\nu },
\ee
with the basic property $u_{\nu }h_{\lambda }^{\nu }\equiv 0$.

By taking the covariant divergence of Eq.~(\ref{em}), we first find
\begin{eqnarray}
\nabla ^{\mu }T_{\mu \nu } &=&\left( \nabla ^{\mu }\rho +\nabla ^{\mu
}p\right) u_{\mu }u_{\nu }+\left( \rho +p\right) \nabla ^{\mu }u_{\mu
}u_{\nu }  \notag \\
&+&\left( \rho +p\right) u_{\mu }\nabla ^{\mu }u_{\nu }+\nabla_\nu p.
\end{eqnarray}%

Therefore Eq.~(\ref{noncons}) takes the form
\begin{align}  \label{em1}
 \nabla ^{\mu }(\rho&+p) u_{\mu }u_{\nu }+\left(
\rho +p\right) \nabla ^{\mu }(u_{\mu }u_{\nu })  \notag \\
&+\nabla _{\nu}p= \frac{%
1}{8\pi +f_{m}}\bigg[\nabla _{\nu }\left(L_{m}f_{m}\right)  \notag \\
&-T_{\mu \nu }\nabla ^{\mu }f_{m}- \frac{1}{2}(f_{T}\nabla _{\nu }T+f_{L}\nabla _{\nu }L_{m})\bigg],
\end{align}
where we have taken into account that $A_\nu=0$‌ in the case of a perfect fluid.
By multiplying Eq.~(\ref{em1}) with $u^{\nu }$, and by considering
the identity $u^{\nu }\nabla ^{\mu }u_{\nu }=0$, we arrive at the energy
balance equation in the $f\left( R,L_{m},T\right) $ gravity theory as given
by
\begin{align}  \label{em2}
\dot{\rho}&+3\left( \rho +p\right) H=\nonumber\\&-\frac{1}{8\pi +f_{m}}\Bigg[(\rho+L_m)\dot{f}_m-\f12f_T(\dot T-2\dot{L}_m)\Bigg],
\end{align}%
where we have denoted $H=(1/3)\nabla ^{\mu }u_{\mu }$, and $\dot{}=u^{\mu
}\nabla _{\mu }=d/d{s}$, respectively. Here $d{s}$ denotes the line element
constructed with the help of the metric $g_{\mu \nu }$, $d{s}^{2}=g_{\mu \nu }dx^{\mu
}dx^{\nu }$. We note that the equality $u^{\mu
}\nabla _{\mu }=d/d{s}$ is only true when we apply it on a scalar function. In the case of tensors, we will substitute the definition with the absolute derivative $u^{\mu
}\nabla _{\mu }=D/D{s}$, as we will see in the following.

The multiplication of Eq.~(\ref{em1}) with $h_{\lambda }^{\nu }$ gives the momentum
balance equation for a perfect fluid in the alternative $f\left(
R,L_{m},T\right) $ gravity theory as
\begin{align}  \label{force}
u^{\mu }\nabla _{\mu }u^{\lambda } &=\frac{d^{2}x^{\lambda }}{ds^{2}}%
+\Gamma _{\mu \nu }^{\lambda }\f{dx^\mu}{ds}\f{dx^\nu}{ds}\nonumber\\&=\frac{1}{(\rho
		+p)(8\pi +f_{m})}\Big[(L_m-p)D^\lambda f_m\nonumber\\&-\f12f_TD^\lambda(T-2L_m)-(8\pi +f_{m})D^\lambda p\Big],
\end{align}
where we have defined $D^\lambda\equiv h^{\nu\lambda}\nabla_\nu$, with $D^{\lambda}$ orthogonal to the vector field $u^\nu$, and lying in the spatial hypersurfaces of constant time.

Alternatively, we can write down the equation of motion of particles as
\begin{equation}
\frac{d^{2}x^{\lambda }}{ds^{2}}+\Gamma _{\mu \nu }^{\lambda }u^{\mu }u^{\nu
}=f^{\lambda },
\end{equation}%
where
\begin{align}  \label{force1}
f^{\lambda } =&\frac{1}{(\rho
	+p)(8\pi +f_{m})}\Big[(L_m-p)D^\lambda f_m\nonumber\\-&\f12f_TD^\lambda(T-2L_m)-(8\pi +f_{m})D^\lambda p\Big],
\end{align}
or, equivalently,
\begin{eqnarray}\label{force1a}
f^{\lambda } &=&\frac{h^{\nu \lambda }}{(\rho +p)(8\pi +f_{m})}\Big[%
(L_{m}-p)\nabla _{\nu }f_{m}  \nonumber  \\
&&-\frac{1}{2}f_{T}\nabla _{\nu }(T-2L_{m})-(8\pi +f_{m})\nabla _{\nu }p\Big],
\end{eqnarray}
is the extra-force generated by the geometry-matter coupling in the $f\left(R,L_m,T\right)$ theory. It is straightforward to see that the extra-force is perpendicular to the four-velocity, $u_{\lambda}f^{\lambda}=0$.

\subsection{The variational principle for the equation of motion of a test particle}

In the following we assume that the extra-force in the $f\left(R,L_m,T\right)$ theory, as given by Eq.~(\ref{force1}%
), can be represented in a formal mathematical way as
\begin{equation}  \label{force2}
f^{\lambda}=-\left( u^{\nu }u^{\lambda }+g^{\nu \lambda
}\right)\nabla _{\nu}\ln \sqrt{Q}=-h^{\nu \lambda}\nabla _{\nu}\ln \sqrt{Q},
\end{equation}
where the expression of $\nabla _{\nu}\ln \sqrt{Q}$ can be obtained after a simple
comparison with Eq.~(\ref{force1a}), once the full set of parameters of the $f\left(R,L_m,T\right)$ theory is
established. If it happens that the extra-force admits such a mathematical representation, the equation of
motion of the massive particles can be derived from a variational principle, which is given by
\begin{equation}
\delta S_{p}=\delta \int L_{p}ds=\delta \int \sqrt{Q}\sqrt{-g_{\mu \nu
}u^{\mu }u^{\nu }}ds=0\,,  \label{actpart}
\end{equation}
where $S_{p}$ and $L_{p}=\sqrt{Q}\sqrt{-g_{\mu \nu }u^{\mu }u^{\nu }}$ are
the action and the Lagrangian density for  massive test bodies moving in a curved geometry described
by the metric $g_{\mu\nu}$. As usual, by $s$ we have denoted the affine parameter along the curve.

To prove this result we begin by writing down  the Euler-Lagrange equations
obtained from the variation of the action~(\ref{actpart}), and which are given by
\begin{equation}
\frac{d}{ds}\left( \frac{\partial L_{p}}{\partial u^{\lambda }}\right) -%
\frac{\partial L_{p}}{\partial x^{\lambda }}=0.
\end{equation}

By considering the mathematical relations
\begin{equation}
\frac{\partial L_{p}}{\partial u^{\lambda }}=\sqrt{Q}u_{\lambda },
\end{equation}
and
\begin{equation}
\frac{\partial L_{p}}{\partial x^{\lambda }}=\frac{1}{2} \sqrt{Q}g_{\mu
\nu,\lambda }u^{\mu }u^{\nu }+\frac{ 1}{2} \frac{Q_{,\lambda }}{\sqrt{Q}},
\end{equation}
respectively, after a simple calculation we obtain the equations of
motion of the particle in the form
\begin{equation}\label{eqmot}
\frac{d^{2}x^{\mu }}{ds^{2}}+\Gamma _{\nu \lambda }^{\mu }u^{\nu }u^{\lambda
}+\left( u^{\mu }u^{\nu }+g^{\mu \nu }\right) \nabla _{\nu }\ln \sqrt{Q}=0.
\end{equation}

By the simple identification of the terms in the equation of motion given by
Eq.~(\ref{eqmot}), and the expression of the extra-force generated by the geometry-matter coupling, as given by Eq.~(\ref%
{force2}), we can immediately obtain the explicit functional form of $\sqrt{Q}$. In the limit $%
\sqrt{Q}\rightarrow 1$, we reobtain the usual general relativistic
equation $\frac{d^{2}x^{\mu }}{ds^{2}}+\Gamma _{\nu \lambda }^{\mu }u^{\nu }u^{\lambda
}=0$, describing geodesic motion in the gravitational field.

\subsection{The Newtonian limit of the equation of motion}

The variational principle~(\ref{actpart}) provides  a very simple and
efficient way for the study of the Newtonian limit of the equations of
motion of the test particles in $f\left( R,T,L_{m}\right) $ gravity. In the
Newtonian limit of the weak gravitational fields and slow motions, one can approximate the metric according to
\begin{equation}
\sqrt{-g_{\mu \nu}u^{\mu }u^{\nu }}ds\approx \sqrt{1+2\phi -\vec{v}^{\,2}}%
dt\approx \left( 1+\phi -\frac{\vec{v}^{\;2}}{2}\right) dt\,,
\end{equation}%
where by $\phi $ we have denoted the Newtonian potential, and $\vec{v}$ is the usual
tridimensional velocity of the fluid, satisfying the condition $\vec{v}^{\,2}/c^2\ll1$. By assuming that in the Newtonian
limit of weak gravitational fields the function $\sqrt{Q}$ can be
represented in the $f\left(R,L_m,T\right)$ theory as,
\begin{equation}\label{41}
\sqrt{Q}=1+U\left( \phi,\rho ,p,T, L_m,f_m, \nabla _{\nu}T,...\right) ,
\end{equation}%
where we assume that in the first order of approximation $U\left( \phi,\rho ,p,T, L_m,f_m, \nabla _{\nu}T,...\right)\ll 1$, the equations of motion of the fluid do follow from the variational principle
\begin{equation}
\delta \int \left[ U\left(\phi,\rho ,p\right) +\phi -\frac{\vec{v}^{\,2}%
}{2}\right] dt=0\,,
\end{equation}%
and are obtained as
\begin{equation}
\vec{a}=-\nabla \phi -\nabla U=\vec{a}_{N}+\vec{a}_{E}\,,
\end{equation}%
where $\vec{a}$ denotes the total acceleration of the system, the term $\vec{a}%
_{N}=-\nabla \phi $ gives the Newtonian gravitational acceleration, while
\begin{equation}
\vec{a}_{E}=-\nabla U\left( \phi,\rho ,p,T, L_m,f_m, \nabla _{\nu}T,...\right),
\end{equation}%
represents the supplementary acceleration induced by the inclusion of the different
forms of the geometry-matter coupling in the action of the gravitational
field of the $f\left(R,L_m,T\right)$ theory.

In the  case of the extra-force generated in the $f\left(R,L_m,T\right)$ gravity, the expression of $Q$ can be obtained from the relation
\begin{align}\label{eq45}
\nabla _{\nu }\ln&\sqrt{Q}=-\frac{1}{(\rho +p)(8\pi +f_{m})}\Bigg\{(L_{m}-p)\nabla _{\nu
}f_{m}\nonumber\\
&-\frac{1}{2}f_{T}\nabla _{\nu }(T-2L_{m})-(8\pi +f_{m})\nabla _{\nu }p\Bigg\}.
\end{align}

The function $\sqrt{Q}$ can always be represented in a closed form obtained by integrating the left-hand side of  Eq.~(\ref{eq45}). But we must point out that generally $Q$ cannot be expressed in a simple mathematical form, and to find its expression some approximate methods must be used.

In order to obtain the explicit expressions of the extra-acceleration induced by the geometry-matter coupling, we analyze the simple case in which the pressure of the gravitating fluid is given as a
function of the density by a linear barotropic equation of state having the standard form $p=w\rho $, where $w$ is a constant satisfying the
condition $w\ll 1$. Hence it follows that we can approximate $\rho +p\approx \rho $, and $T=-\rho
+3p\approx -\rho$, respectively.  Hence in this approximation  $T$ is a function of the density only. Furthermore,  we assume that the function $f\left(R,T,L_m\right)$ can be represented as $f\left(R,T,L_m\right)=R+\epsilon \Psi \left(L_m,T\right)$, where $\epsilon$ is a small parameter. We also adopt for the matter Lagrangian the expression $L_m=-\rho$. Hence the gravitational Lagrangian density becomes a function of the Ricci scalar and of the matter density only, $f\left(R,T,L_m\right)=R+\epsilon \Psi \left(\rho \right)$.

Now we can expand $\Psi (\rho)$ near a fixed value $\rho_0$ of the
matter density, thus obtaining in the first order of approximation
\bea
\Psi\left(\rho \right)&\approx& \Psi\left(\rho _0\right)
+\left(\rho -\rho _0\right)\left.\frac{d\Psi (\rho)}{d\rho}
\right|_{\rho=\rho_0}+... \nonumber\\
&=&a_0+ b_0\left(\rho -\rho_0\right)+...,
\eea
where we have denoted $a_0=\Psi\left(\rho _0\right)$ and
$b_0=\left[d\Psi (\rho)/d\rho \right]|_{\rho=\rho_0}$, respectively. Hence we immediately find $f_T=\epsilon b_0$, $f_L=-\epsilon b_0$, and $f_m=f_T+f_L/2=\epsilon b_0/2$, respectively.

Then, with the use of the above approximations,  Eq.~(\ref{eq45}) giving the
expression of $\sqrt{Q}$ takes the form
\begin{equation}\label{Q1}
\nabla _{\nu }\ln \sqrt{Q}\approx \left(w-\frac{\epsilon b_0/2}{8\pi+\epsilon b_0/2}\right)\frac{1}{\rho}\nabla _{\nu} \rho,
\end{equation}
giving
\begin{equation}\label{eq48}
\sqrt{Q}\left(\rho \right)\approx \left(\frac{\rho }{\rho _0}\right)^\alpha,
\end{equation}
where $\alpha =w-\left(\epsilon b_02\right)/\left(8\pi +\epsilon b_0/2\right)$.  Eq.~(\ref{eq48}) also maintains its validity for a fluid satisfying a linear barotropic equation of state of the form $p=\left(\gamma -1\right)\rho $, $\gamma ={\rm constant}$, with $(\gamma -1)$ not necessarily very small. Hence an equation of the form (\ref{eq48}) can be applied for the description of both non-relativistic and the extreme relativistic regimes.

By using the mathematical relation $x^{\alpha }=\exp (\alpha \ln x)\approx 1+\alpha
\ln x$,
we can approximate $\sqrt{Q}\left(\rho \right)$ given by Eq.~(\ref{eq48}) as
\be
\sqrt{Q}\left(\rho \right)\approx 1+\alpha \ln \frac{\rho }{\rho _0}=1+U(\rho),
\ee
where
\be
U\left(\rho \right)= \alpha \ln \frac{\rho}{\rho _0}.
\ee

Hence in the first order of approximation in $f\left(R,L_m,T\right)$ gravity the equations of motion of the gravitating
hydrodynamic type system can be obtained from the variational principle
\begin{equation}
\delta \int \left[ 1+\alpha \ln \frac{\rho}{\rho _0}
+\phi -\frac{\vec{v}^{\,2}}{2}\right] dt=0\, ,
\end{equation}
and they are given by
\begin{equation}
\vec{a}=-\nabla \phi -\nabla U\left(\rho
\right)=\vec{a}_{N}+\vec{a}_{E}\, ,
\end{equation}
with the extra-acceleration of the fluid given by
\begin{equation}
\vec{a}_{E}(\rho)=\alpha \nabla \ln \frac{\rho }{\rho_0} =\alpha \frac{1}{\rho}\nabla \rho.
\end{equation}
In the present approach and within the framework of the adopted approximations the supplementary acceleration induced by the modification of the
action of the gravitational field due to the geometry-matter coupling is a function of the matter density only.

\subsection{The modified Poisson equation}

In this Subsection we will obtain the modified Poisson equation in $%
f(R,L_m,T)$ gravity. In order to find it we consider the first order
approximation of the function $f\left(R,L_m,T\right)$, which can be written
with the help of a Taylor series expansion in the vicinity of a fixed point $%
\left(R_0,T_0,L_{m0}\right)$ as
\begin{align}
f&\left(R,T,L_m\right)\approx \lambda +\alpha R+\beta _0T+\gamma _0L_m\nonumber\\
&+\mathcal{O}\left(R-R_0\right)^2+\mathcal{O}\left(L-L_{m0}\right)^2+\mathcal{O}\left(T-T_0\right)^2,
\end{align}
where we have denoted
\begin{eqnarray}
&&\alpha =f_R\left(R_0,T_0,L_{m0}\right), \beta
_0=f_T\left(R_0,T_0,L_{m0}\right),   \\
&&\gamma _0=f_L\left(R_0,T_0,L_{m0}\right),   \\
&&\lambda =f\left(R_0,T_0,L_{m0}\right)- \alpha R_0-\beta T_0-\gamma L_{m0}.
\end{eqnarray}

Therefore, in
this approximation, the field equations (\ref{eq2}) for a perfect fluid with
energy-momentum tensor given by Eq.~(\ref{em}) become
\begin{equation}  \label{43}
R_{\mu \nu}-\frac{1}{2}Rg_{\mu \nu}=\Lambda g_{\mu \nu}+\beta Tg_{\mu \nu}+8\pi G_{eff}T_{\mu \nu},
\end{equation}
where we have denoted
\begin{equation}
\beta =\frac{\beta _0}{2\alpha}, \gamma =\frac{\gamma _0}{2\alpha}, \Lambda =%
\frac{\lambda}{2\alpha}, G_{eff}=\left[\frac{1}{\alpha}+\frac{\gamma +2\beta%
}{8\pi}\right].
\end{equation}

Equivalently, Eq.~(\ref{43}) can be reformulated as
\begin{equation}  \label{44}
R_{\mu \nu}=-\Lambda g_{\mu \nu}-\beta Tg_{\mu\nu}+8\pi
G_{eff}\left(T_{\mu \nu}-\frac{1}{2}g_{\mu \nu}T\right).
\end{equation}

In the Newtonian limit one can write the metric as
\begin{align}
ds^2=-(1+2\phi(\vec{x}))dt^2+d\vec{x}^2,
\end{align}
where $\phi$ is the Newtonian potential. In the following we assume that the
macroscopic motion takes place at small velocities, and consequently in the four-velocity we can keep
only the time component, and neglect all the space components with a very good approximation. Therefore we have
$u^0\approx  -1$, and $u^i\approx 0$, $i=1,2,3$.

We will apply the same
approximation in the gravitational field equations, thus keeping only their time component. A similar approximation is used for the components of the energy-momentum tensor. Therefore, we can
approximate $T_{\mu \nu}$ as $T_0^0\approx-\rho $, and $T=-\rho$,
respectively. By taking into account the explicit expression of the metric tensor, we obtain
$\Gamma _{00}^{\alpha}\approx -(1/2)g^{\alpha \beta}\partial g_{00}/\partial
x^{\beta}=\partial \phi /\partial x^{\alpha}$, giving $R_0^0=-\Delta \phi$. Then Eq.~(%
\ref{44}) immediately gives
\begin{equation}
\Delta \phi =4\pi G_{New}\rho +\Lambda,
\end{equation}
with $G_{New}=G_{eff}-\beta/4\pi$.
Hence, we have obtained the generalized Poisson equation in $%
f\left(R,L_m,T\right)$ gravity theory, which, in the first order of approximation, determines a slight modification of
the gravitational constant, and induces a cosmological term.

\section{The Dolgov-Kawasaki instability}\label{4}

In this Section we will obtain the Dolgov-Kawasaki criterion \cite{DK} in $f\left(R,L_m,T\right)$ gravity theory. Let's consider the dynamics of a small celestial body like, for example, the Sun. In this case, the gravitational force is very small, and the metric is very close to the flat one. However, the curvature of the space-time is still non-zero, and the geometry is curved. Suppose that there exist a solution of the field equations corresponding to constant values of the variables of the theory, $R_0$, $T_0$ and $L_{0}$, respectively. Let us now perturb the curvature, the matter energy-momentum tensor, and the matter Lagrangian  around their background values as
\begin{align}
R&=R_0+R_1,\\
T&=T_0+T_1,\\
L_m&=L_{0}+L_1.
\end{align}
Also, we can expand the tensor $\tau_{\mu\nu}$ as $\tau_{\mu\nu}=\tau_{0\mu\nu}+\tau_{1\mu\nu}$.
In order to be consistent with the Solar System observations, in addition to the above relations, we assume that the function $f$ can be expressed as
\begin{align}
f(R,L_m,T)=R+\epsilon \Phi\left(R,L_m,T\right),
\end{align}
where $\epsilon$ is a small constant, with its value set by Solar System tests, and $\Phi\left(R,L_m,T\right)$ is an arbitrary function.

Let us start with the trace of Einstein's field equation \eqref{tr}. To the first order in perturbations, we have
\be
f(R,L_m,T)=R_0+\epsilon\Phi(0)+(1+\epsilon\Phi_R(0))R_1+H_1,
\ee
where $H_1$ contains all first order contributions of the matter field, i.e. $T_1$ and $L_1$. One can also obtain
\begin{align}
f_R&=1+\epsilon\Phi_R(0)+\epsilon\Phi_{RR}R_1+H_{R1},\\
f_T&=\epsilon\Phi_T(0)+\epsilon\Phi_{RT}(0)R_1+H_{T1},\\
f_L&=\epsilon\Phi_L(0)+\epsilon\Phi_{RL}(0)R_1+H_{L1}.
\end{align}

The higher order derivatives of the function $f$ can be  easily written as above. We also have
\be
\Box f_{1R}=\epsilon\Phi_{RR}(0)\Box R_1+\epsilon\Phi_{RT}(0)\Box T_1+\epsilon\Phi_{RL}(0)\Box L_1.
\ee
By using the above approximations, it follows that the d'Alembertian can be expanded in the Minkowski space as
\begin{align}
\Box  R_1=-\ddot{R}_1+\nabla^2 R_1.
\end{align}
The resulting trace equation \eqref{tr} at first order can be written as
\begin{align}
\ddot{R}_1-\nabla^2 R_1+m_{eff}^2 R_1=V_{eff},
\end{align}
where we have defined the effective mass, and effective potential, respectively, as
\begin{align}\label{effmass}
m_{eff}^2&=\f{1}{3\epsilon\Phi_RR(0)}\Bigg(1+\epsilon\Phi_R(0)-\epsilon\Phi_{RR}(0)R_0\nonumber\\&-2\epsilon\Phi_{RL}(0)L_0-4\epsilon\Phi_{TR}L_0+\f12\epsilon\Phi_{LR}(0)T_0\nonumber\\&+\epsilon\Phi_{TR}(0)T_0+\epsilon\Phi_TR(0)\tau_0\Bigg),
\end{align}
and
\begin{align}
V_{eff}^2&=\f{1}{3\epsilon\Phi_RR(0)}\Bigg(2H_1-H_{R1}R_0-2(H_{L1}+2H_{T1})L_0\nonumber\\&-2\epsilon(\Phi_{L}(0)+2\Phi_T(0))-3\epsilon\Phi_{RT}(0)\Box T_1\nonumber\\&-3\epsilon\Phi_{RT}(0)\Box L_1+\f12\big((H_{L1}+2H_{T1})T_0+16\pi T_1\nonumber\\&+\epsilon(\Phi_L(0)+2\Phi_T(0))T_1\big)+H_{T1}\tau_0+\epsilon\Phi_T(0)\tau_1\Bigg).
\end{align}

In order to have a stable perturbation, and to avoid tachyonic instability, one must impose the condition  $m_{eff}^2\geq0$. However, the mass term \eqref{effmass} is dominated by its first term, which gives us the constraint $\Phi_{RR}(0)\geq0$. Noting that $\Phi_{RR}(0)=f_{RR}(0)$, we finally obtain the Dolgov-Kawasaki criterion in $f\left(R,L_m,T\right)$ gravity as
\begin{align}
f_{RR}(0)\geq0.
\end{align}
The same condition applies in the special cases of $f(R)$, $f(R,T)$ and $f(R,L_m)$ theories, which were proven elsewhere.

\section{Cosmological consequences of $f\left(R,L_m,T\right)$ gravity}\label{5}

Let us assume that the Universe can be described by the flat, homogenous and isotropic Friedmann-Lemaitre-Robertson-Walker (FLRW) metric of
the form
\begin{equation}
ds^{2}=-dt^{2}+a(t)^{2}(dx^{2}+dy^{2}+dz^{2}),
\end{equation}%
where $a(t)$ is the scale factor. In the following we denote by $H=\dot{a}/a$
the Hubble function, describing the rate of expansion of the Universe.
The accelerating/decelerating nature of the cosmological expansion
can be characterized with the help of the deceleration parameter $q$, defined as%
\begin{equation}
q=\frac{d}{dt}\left( \frac{1}{H}\right) -1.
\end{equation}

Also, let us assume that the Universe is filled with a perfect fluid. We adopt for the
Lagrangian density of the cosmic (baryonic) matter the expression $L_{m}=-\rho $. In the comoving frame the non-zero components of the  energy-momentum tensor are given by
\begin{equation}
T_{\nu }^{\mu }=\mathrm{diag}\left( -\rho ,p,p,p\right) .
\end{equation}

Moreover, we impose on the function $f$ the condition $f_{R}\neq 0$, a condition that
is assumed to be valid for all times.

\subsection{Generalized Friedmann equations}

In the case of
\begin{it}
	the traceless representation of the $f\left( R,L_{m},T\right)
	$ gravitational theory,
\end{it}
 the traceless equation \eqref{13} and the trace equation \eqref{tr}  reduce for the FLRW metric to the evolution
equations (generalized Friedmann equations)
\begin{align}\label{traceless}
-\dot{H}=\frac{1}{2f_{R}}%
\left( 8\pi +f_{m}\right) \left( \rho +p\right)
+\frac{1}{2f_{R}}\left( \ddot{f}_{R}-H \dot{f}_{R}\right),
\end{align}
and
\begin{align}\label{trace}
\dot H&+2H^2=\f{1}{3f_R}(f-2L_mf_m)\nonumber\\&-\frac{1}{6f_{R}}%
\left( 8\pi +f_{m}\right) \left( \rho -3p\right)+\frac{1}{2f_{R}}\left( \ddot{f}_{R}+3H \dot{f}_{R}\right).
\end{align}
In order to obtain a consistent solution it is usually necessary to solve both cosmological equations above. This is due to the fact that the original set of Einstein equations are decomposed into a traceless equation, and the trace equation, with $L_m$ appearing only in the second equation.

To close the system of field equations we must add to cosmological system the energy balance equation (\ref{conssim}), which
gives the time evolution of the density as
\begin{align}\label{balance}
\dot{\rho}+4H\left( \rho +p\right) &=-\dot{p}-\frac{1}{3\left( 8\pi
+f_{m}\right) }Rf_{R}\frac{d}{dt}\ln \frac{R}{f_{R}}  \notag \\
&-\left( \rho +p\right) \frac{d}{dt}\ln \left( 8\pi +f_{m}\right)   \notag
\\
&-\frac{1}{\left( 8\pi +f_{m}\right) }\frac{d}{dt}\left[ \frac{1}{a^{3}}%
\frac{d}{dt}\left( a^{3}\frac{df_{R}}{dt}\right) \right] .
\end{align}
The deceleration parameter can be obtained in a general form as

\begin{align}
q=\frac{1}{2f_{R}H^{2}}\left[ \left( 8\pi +f_{m}\right) \left( \rho
+p\right) +\ddot{f}_{R}-H\dot{f}_{R} \right]
-1.
\end{align}

Eq.~(\ref{traceless}) immediately gives the main criterion for the existence of an
asymptotic de Sitter type evolution with $H=\mathrm{constant}=H_{0}$ in $f\left(R,L_m,T\right)$ gravity as
\begin{equation}
\left( 8\pi +f_{m}\right) \left( \rho +p\right) +\left( \ddot{f}_{R}-H_0\dot{f}_{R}\right) =0.  \label{accel}
\end{equation}

Assuming that matter and geometry are minimally coupled  $f\left( R,L_{m},T\right) =R+g\left(
L_{m},T\right) $, the condition for accelerated de Sitter type expansion
reduces to
\begin{equation}
8\pi +\frac{\partial g\left( L_m,T\right) }{\partial T}+\frac{1}{2}\frac{%
\partial g\left(L_{m},T\right) }{\partial L_{m}}=0,
\end{equation}%
representing a linear first order partial differential equation, with the
general solution given by%
\begin{equation}
g\left( T,L_{m}\right) =-8\pi \big[\alpha T+2(1-\alpha)L_m\big]+C\left(L_{m}- \frac{1}{2}T
\right) ,
\end{equation}%
where $\alpha$ is an arbitrary constant and $C$ is an arbitrary function of the variable $\left( 2L_{m}-T\right) /2$. In the simple case of an additive structure of the gravitational Lagrangian of the form $f\left(R,L_m,T\right)=R+\alpha L_m+\beta T$,  the condition for de Sitter type expansion is given by $8\pi +\alpha /2+\beta =0$.

The general condition for an accelerating expansion of the Universe, with $q<0$, can be formulated as
\be\label{eq87}
\ddot{f}_R-H\dot{f}_R+\left(8\pi+f_m\right)\left(\rho +p\right)<2f_RH^2.
\ee

The condition given by Eq.~(\ref{eq87}) depends on the Hubble parameter and the time/redshift derivative of the function $f\left(R,L_m,T\right)$. Hence, once the functional form of $f\left(R,L_m,T\right)$ is specified, the sign of the above condition changes for different time/redshift intervals, thus allowing for the existence of both decelerating and accelerating phases during the cosmological evolution.

\subsection{The multiplicative case: $f\left(R,L_m,T\right)=R+\alpha TL_m+\beta$, with $L_m=-\rho$}

In order to build some explicit cosmological models in $f\left(R,L_m,T\right)$ theory we will consider some explicit expressions for $f$. Let us consider first the choice
\be
f\left(R,L_m,T\right)=R+\alpha TL_m+\beta,
\ee
for the Lagrangian density.

\subsubsection{The radiation dominated Universe}

In the case of radiation dominated Universe, we have $p=\rho/3$, and the evolution equations \eqref{traceless}, \eqref{trace} and \eqref{balance} can be simplified to
\begin{align}\label{rad1}
& 3 \dot{H}-2   (\alpha  \rho -8 \pi )\rho=0,\nonumber\\
& 6 H^2+3 \dot{H}-2 \alpha  \rho ^2+\beta =0,
\end{align}
and
\begin{align}
2 \pi \dot{\rho}- (\alpha  \rho-8 \pi  )H\rho=0,
\end{align}
respectively. The above equations have an exact solution representing the radiation dominated Universe. Eliminating $\dot{H}$ in Eqs. \eqref{rad1} one can easily obtain
\begin{align}
H=\sqrt{\frac{8 \pi  \rho }{3}-\frac{\beta }{6}},
\end{align}
substituting the above equation into the field equations, we have
\begin{align}
	t-t_0=-&\sqrt{\f{3}{2\beta}}\,\tan^{-1}\left(\sqrt{\f{16\pi}{\beta}\rho-1}\right)\nonumber\\&-\sqrt{\f{3\alpha}{2\mathcal{A}}}\,\tanh^{-1}\left(\sqrt{\f{16\pi\alpha\rho-\alpha\beta}{\mathcal{A}}}\right),
\end{align}
where $\mathcal{A}=128\pi^2-\alpha\beta$.

Also, in the case $\alpha>0$, we have an accelerating solution with the Hubble parameter
\begin{align}
	H=H_0=\sqrt{\f{\mathcal{A}}{6\alpha}},
\end{align}
with energy density $\rho=8\pi/\alpha$.

\subsubsection{The dust Universe}

In the case of dust dominated Universe, one has the evolution equations
\begin{align}\label{80}
4\dot{H}+ \left(16\pi-3 \alpha
\rho\right)\rho =0,
\end{align}
\begin{align}\label{81}
4 \left(3 \dot{H}+6 H^2+\beta \right)-(5 \alpha  \rho +16 \pi )\rho=0,
\end{align}
and
\begin{align}
2 (8 \pi -\alpha  \rho ) \dot{\rho}+3  (16 \pi -3 \alpha  \rho
) H\rho=0,
\end{align}
respectively.

By using Eqs.~(\ref{80}) and (\ref{81}), we obtain
\be\label{84}
H=\frac{\sqrt{ (16 \pi -\alpha  \rho )\,\rho-\beta }}{\sqrt{6}}.
\ee

Substituting the above relation for the Hubble function in the field equation \eqref{81} we find
\begin{align}
t-t_0=\sqrt{\f{2\alpha}{3\mathcal{B}}}&\ln\left[\f{16\pi-3\alpha\rho}{\sqrt{\alpha}\left(128\pi^2-3\alpha\beta +8\alpha \rho+\sqrt{\alpha \mathcal{B}\mathcal{C}}\right)}\right]\nonumber\\&
+\sqrt{\f{2}{-3\beta}}\ln\left[\f{4(8\pi\rho-\beta)}{\sqrt{-\beta}}+\sqrt{\mathcal{C}}\right],
\end{align}
where we have defined
\begin{align}
	\mathcal{B}=512\pi^2-9\alpha\beta, \quad\mathcal{C}=16\pi\rho-\alpha \rho^2-\beta.
\end{align}

 This model has also a de Sitter type expansionary phase with $H=H_0={\rm constant}$, for $H_0$ given by
 \begin{align}
 H_0=\sqrt{\f{\mathcal{B}}{54\alpha}}, \quad \rho=\f{16\pi}{3\alpha}.
 \end{align}

Hence in this model the de Sitter type phase is triggered by the generalized geometry-matter coupling, and takes place in the presence of ordinary matter only.

\subsubsection{Numerical results}

Now, we consider the cosmological evolution of the Universe in the present model by assuming it is filled with a mixture of dust and radiation, with the total energy density given by
\begin{align*}
\rho=\rho_r+\rho_m.
\end{align*}

The equation of state of the radiation is $p=p_r=\rho_r/3$.  In order to simplify the mathematical formalism we introduce a set of dimensionless parameters defined as
\begin{align}
H&=H_0 h,\quad \rho_i= 6\kappa^2 H_0^2 \Omega_i,\quad i=r,m\nonumber\\
\bar{\alpha}&=  6\kappa^4 H_0^2 \,\alpha,\quad \beta =3\,H_0^2 \,\bar{\beta},
\end{align}
where $H_0$ is the current value of the  Hubble function.

In terms of the redshift
$$
1+z=\f1a,
$$
the equations of motion \eqref{traceless} and \eqref{trace} take the form
\begin{align}\label{trls1ro}
2  (1+z)&
h\,h'-3 \Omega _m-4 \Omega _r\nonumber\\&+\bar{\alpha } \left(18 \Omega _m \Omega _r+9 \Omega _m^2+8 \Omega _r^2\right)=0,
\end{align}
and
\begin{align}\label{tr1ro}
2  (1+z)& h\,h'-4 h^2+\Omega _m-2 \bar{\beta }\nonumber\\&+\bar{\alpha } \left(14 \,\Omega _m \Omega _r+5 \Omega _m^2+8 \Omega
_r^2\right)=0,
\end{align}
respectively, where the prime denotes the derivative with respect to the redshift.

One can write the energy balance equation \eqref{balance} as two coupled generalized conservation equations for the dust and the radiation separately as
\begin{align}\label{consm1ro}
3 & \left[\bar{\alpha } \left(3 \Omega _m+2 \Omega _r\right)-1\right]\,\Omega _m\nonumber\\&-(1+z)
\left[\bar{\alpha } \left(2 \Omega _m+\Omega _r\right)-1\right]\Omega _m'=0,
\end{align}
and
\begin{align}\label{consr1ro}
4 \left[\bar{\alpha } \left(3 \Omega _m+2 \Omega _r\right)-1\right]\,\Omega _r -(1+z) \left(\bar{\alpha }
\Omega _m-1\right)\Omega _r' =0,
\end{align}
respectively.

It should be mentioned that the sum of above equations is equal to the original conservation equation \eqref{balance}, and are written in a way that in the limit $\bar\alpha=0$, they reduce to the standard conservation equations for dust and radiation.

 For the initial (present day) values of the cosmological parameters we adopt the values $h(0)=1$, $\Omega_{m0}=0.305$ and $\Omega_{r0}=0.53\times10^{-4}$, respectively. Evaluating Eqs.~\eqref{trls1ro} and \eqref{tr1ro} at the present time $(z=0)$ one can find the numerical values of the constant $\bar{\beta}$ in terms of $\bar{\alpha}$ as
 \begin{align}
 \bar{\beta}=-2+2\left(\Omega_{r0}+\Omega_{m0}\right)\left(1-\bar{\alpha}\,\Omega_{m0}\right).
 \end{align}

 The redshift dependence of the deceleration parameter is given by
 \begin{align*}
 q=-1+(1+z)\f{h'}{h},
 \end{align*}

  Using Eqs.~\eqref{trls1ro} and \eqref{tr1ro} one can obtain the present value of the deceleration parameter $q_0$  as
 \begin{align}
 q_0=-1+\f16 \left(3\Omega_{m0}+4\Omega_{r0}\right)\left(1-3\bar{\alpha }\Omega_{m0}-2\bar{\alpha }\Omega_{r0}\right).
 \end{align}

\begin{figure*}[htbp]
	\includegraphics[scale=0.95]{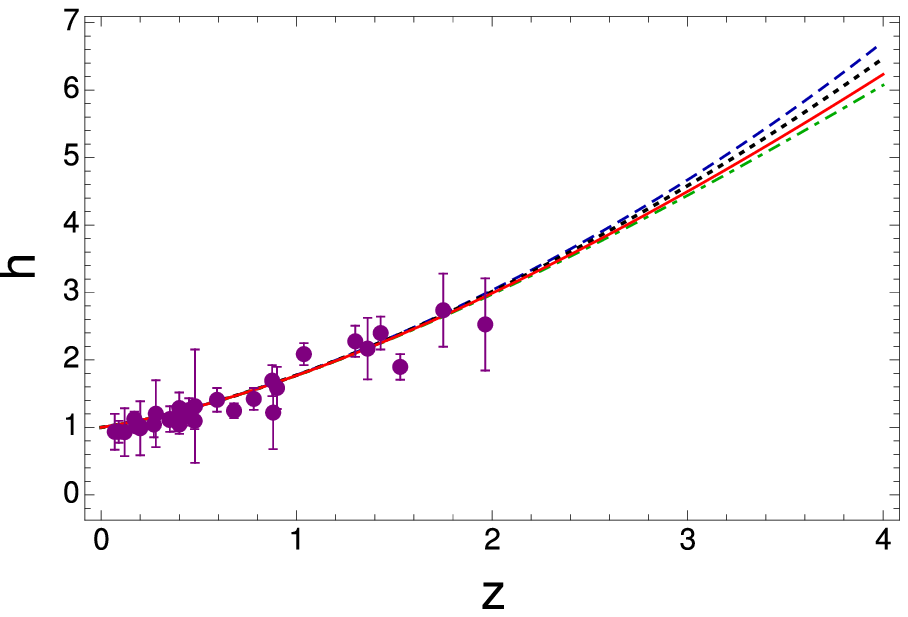}\hspace{.4cm}
	\includegraphics[scale=0.95]{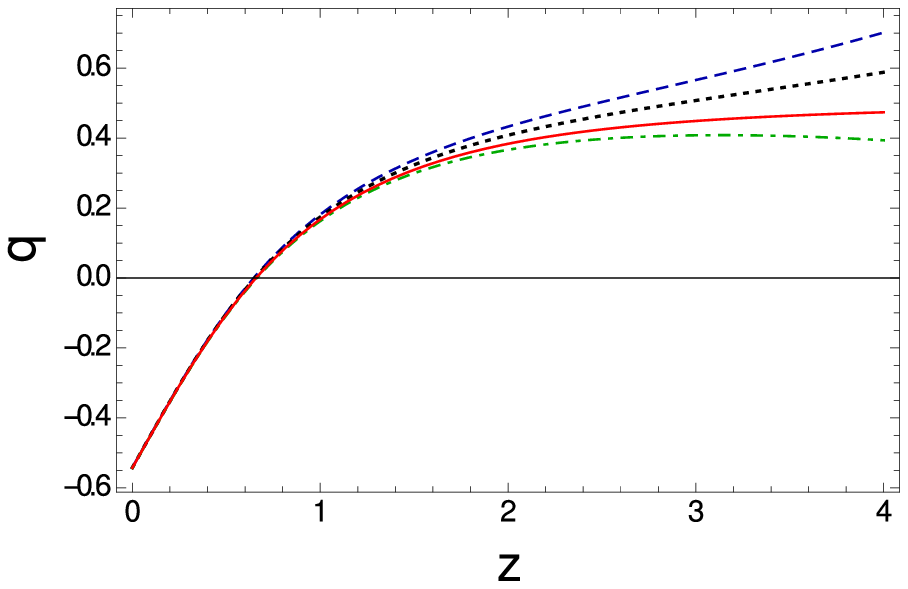}
	\caption{The Hubble function (left panel) and the deceleration parameter (right panel) as a function of the redshift $z$ for the multiplicative model of the $f\left(R,L_m,T\right)$ theory with $L_m=-\rho$ for three different values of the constant $\bar{\alpha }$: $\bar{\alpha }=-0.002$ (dashed curve),   $\bar{\alpha }=-0.001$ (dotted curve) and $\bar{\alpha }=0.0007$ (dot-dashed curve). For the Hubble parameter, we have also plotted the observational data together with their errors \cite{hubble}. The $\Lambda$CDM curve is depicted as a red solid curve.\label{hq1ro}}
\end{figure*}

 The behaviors of the Hubble function and of the deceleration parameter are shown in Fig.~\ref{hq1ro} for three different values of $\bar{\alpha}$.
As one can see from the left panel of Fig.~\ref{hq1ro}, the Hubble function is a monotonically increasing function of $z$, and up to the redshift $z\approx 2$, its evolution is weakly dependent of the numerical values of the model parameter $\bar{\alpha }$. For larger values of the redshift,  significant differences between the model prediction and the observations do appear for some $\bar{\alpha }$ values. However,  $\bar{\alpha }=0.0007$ reproduces well the observed behavior of $H(z)$ up to at least a redshift of $z\approx 4$.
In order to compare the present model with the $\Lambda$CDM theory, let us see the differences (in \%) between the two models at some specific redshifts. For $\bar{\alpha}=-0.002$,  $h$ differs from the $\Lambda$CDM model at redshifts $z=1$ and $z=3$ by $0.35\%$, and $3.8\%$ respectively. Also for $\bar{\alpha}=0.007$ for the deviations of $h$ from the $\Lambda$CDM theory at $z=1$ and $z=3$ one obtains $0.12\%$ and $1.30\%$,  respectively.

 The variation of the deceleration parameter of the model, is represented in the right panel of Fig.~\ref{hq1ro}.
The Universe enters into an accelerating expansionary phase at $z\approx 0.5$, and in the range $z\in(0,1)$ the cosmological evolution is weakly dependent of the numerical values of the model parameters. However, for larger redshifts, the variation of $q$ shows large deviations from the predictions of the $\Lambda$CDM model, with $\bar{\alpha }=0.0007$ giving the closest description to the predictions of standard general relativistic cosmology.

\subsection{The multiplicative case: $f(R,L_m,T)=R+\alpha TL_m$, with $L_m=p$}

It is well-known that there are two choices for the perfect fluid matter Lagrangian, i.e. $L_m=-\rho$ and $L_m=p$. These two choices lead to the same matter energy-momentum tensor in the case of Einstein's gravity.‌‌ However, in the case of modified theories of gravity, these two distinct possibilities lead to different dynamical evolutions of the Universe.

\subsubsection{The radiation dominated Universe}

Now, let us consider the case $L_m=p$. In the case of radiation with $p=\f13\rho$, the evolution equations \eqref{traceless}-\eqref{balance} can be written as
\begin{align}
&3 \dot{H}+\f23 (\alpha  \rho +24\pi ) \rho =0,\nonumber\\
&3 \dot{H}+6 H^2+\beta -\f29 \alpha  \rho ^2=0,
\end{align}
and
\begin{align}
\f23\left(9\pi+\alpha \rho\right)  \dot{\rho}+(24\pi +\alpha  \rho )H \rho  =0.
\end{align}

As in the case of $L_m=-\rho$, the above set of equations admit the radiation dominated solution with
\begin{align}
H=\sqrt{\f{\mathcal{D}}{54}},
\end{align}
and
\begin{align}
&t-t_0=\sqrt{\f{-3}{8\beta}}\,\ln \left[\f{16}{3\rho}\left(\f{3(8\pi\rho-\beta)}{\sqrt{-\beta}}+\sqrt{\mathcal{D}}\right)\right]
-\sqrt{\f{25\alpha}{24\mathcal{A}}}\nonumber\\&\times\ln\bigg[\f{\sqrt{\alpha}\left(3\alpha\beta +576\pi^2+40\pi\alpha\rho\right)-\alpha\sqrt{\mathcal{A}\mathcal{D}}}{\sqrt{\mathcal{A}}(\alpha\rho+24\pi)}\bigg],
\end{align}
where we have defined
$\mathcal{D}=8\alpha \rho^2+144\pi\rho-9\beta$.

\subsubsection{The dust Universe}

In the case of matter dominated Universe with $p=0$, the cosmological evolution equations can be written as
\begin{align}\label{eq2dU}
4 \dot{H}+  (16 \pi -\alpha \, \rho )\rho=0,
\end{align}
\begin{align}\label{eq1dU}
4 \left(3 \dot{H} +6 H^2+\beta\right)+(\alpha  \rho -16 \pi ) \rho=0,
\end{align}
and
\begin{align}
2 (8 \pi -\alpha  \rho ) \dot{\rho}+3  (16\pi -3 \alpha  \rho ) H \rho=0,
\end{align}
respectively.

By subtracting Eqs.~(\ref{eq2dU}) and (\ref{eq1dU}) we obtain  $H(t)$ as
\be
H(t)=\frac{\sqrt{ (16 \pi -\alpha  \rho )\,\rho -\beta }}{\sqrt{6}}.
\ee

Now, by using the expression of the Hubble function in terms of the energy density, and using the field equations, we obtain
\begin{align}
	t-t_0=\int \left(\frac{\sqrt{\frac{2}{3}} (16 \pi -2 \alpha  \rho )}{\rho
		(\alpha  \rho -16 \pi ) \sqrt{16 \pi \rho-\alpha  \rho ^2-\beta  }}\right)d\rho.
\end{align}

\subsubsection{Numerical results}

To describe the general evolution of the Universe we assume again that $\rho=\rho_m+\rho_r$.
The generalized Friedmann  equations in terms of redshift are
\begin{align}\label{pp1}
6  (1+z)&h\,
h'-9 \Omega _m-12 \Omega _r\nonumber\\&+\bar{\alpha } \left(6 \Omega _m \Omega _r+9 \Omega _m^2-8 \Omega _r^2\right)=0,
\end{align}
and
\begin{align}\label{pp2}
18 (1+z)&h\, h'-36 h^2-18 \bar{\beta }\nonumber\\&+\bar{\alpha } \left(6 \Omega _m \Omega _r-9 \Omega _m^2+8 \Omega
_r^2\right)+9 \Omega _m,
\end{align}
respectively. The generalized conservation Eq.\eqref{balance} for dust and radiation takes the form
\begin{align}
3  &\left[\bar{\alpha }\left(3 \Omega _m-2 \Omega _r\right)-3\right]\Omega _m\nonumber\\&\quad-(1+z)
\left[\bar{\alpha } \left(6 \Omega _m+\Omega _r\right)-3\right]\Omega _m' =0,
\end{align}
and
\begin{align}
12&  \left[\bar{\alpha } \left(2 \Omega _r-3 \Omega _m\right)+3\right]\Omega _r\nonumber\\&\quad+(1+z)
 \left[\bar{\alpha } \left(3 \Omega _m-16 \Omega _r\right)-9\right]\Omega _r'=0,
\end{align}
respectively. For this case, with the use of Eqs.~\eqref{pp1} and \eqref{pp2}, the constant parameter $\bar{\beta }$ can be obtained as
\begin{align}
\bar{\beta}=\f29\big[&-9+3(3-\bar{\alpha}\Omega_{r0}-3\bar{\alpha}\Omega_{m0})\Omega_{m0}\nonumber\\&+(9+8\bar{\alpha}\Omega_{r0})\Omega_{r0}\big].
\end{align}

The current value for the deceleration parameter is given by
\begin{align}
q_0=-1+\f16\left(3\Omega_{m0}+4\Omega_{r0}\right)\left[3-\bar{\alpha}\Omega_{m0}+2\bar{\alpha}\Omega_{r0}\right].
\end{align}

\begin{figure*}[htbp]
	\includegraphics[scale=0.95]{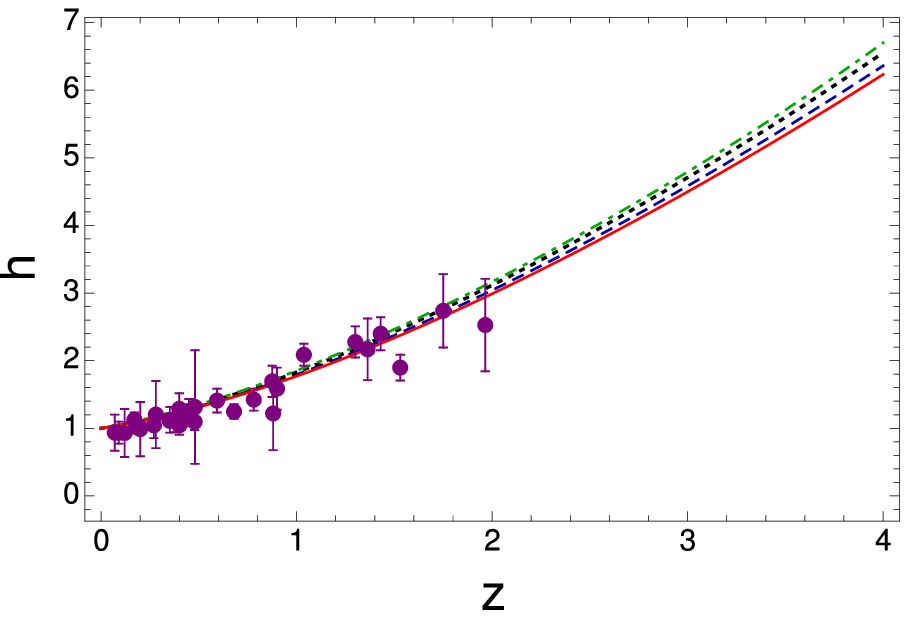}\hspace{.4cm}
	\includegraphics[scale=0.95]{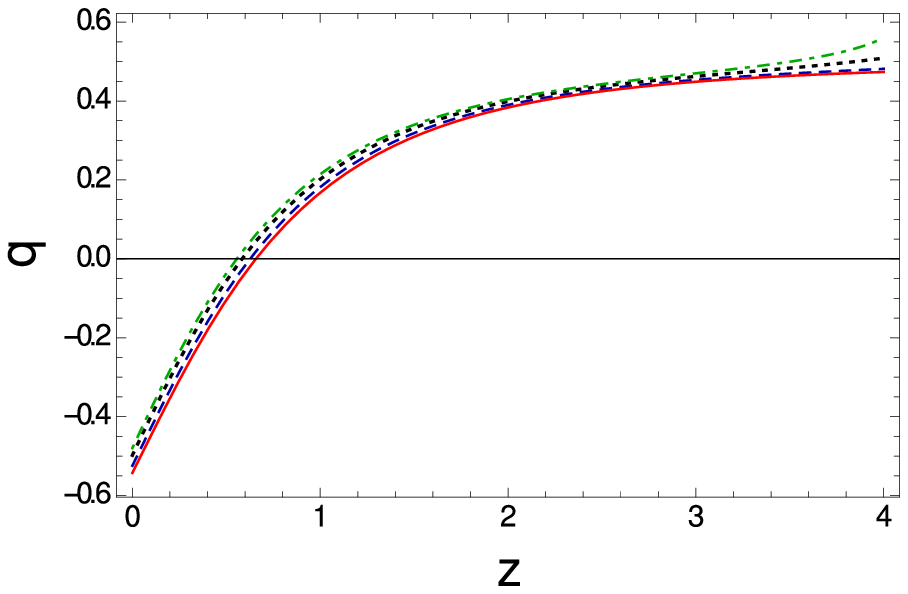}
	\caption{The Hubble function (left panel) and the deceleration parameter (right panel) as a function of the redshift $z$ for the multiplicative model of the $f\left(R,L_m,T\right)$ theory with $L_m=p$ for three different values of the constant $\bar{\alpha }$: $\bar{\alpha }=-0.13$ (dashed curve), $\bar{\alpha }=-0.32$ (dotted curve), and $\bar{\alpha }=-0.45$ (dot-dashed curve), respectively. For the Hubble function, we have also plotted the observational data together with their errors \cite{hubble}. The $\Lambda$CDM curve is depicted as a red solid curve.\label{hq1p}}
\end{figure*}

 The variations of the Hubble function and of the deceleration parameter are shown, as functions of the redshift, in Figs.~\ref{hq1p}.

As one can see from the left panel of Fig.~\ref{hq1p}, up to a redshift of $z\approx 1$, the present model gives a good description of the observational data for $h(z)$, and its predictions almost coincide with the $\Lambda$CDM model. The dependence on the model parameter $\bar{\alpha}$ is weak in the low redshift range. However, for $z>1$, the model behavior is  influenced by the numerical values of $\bar{\alpha}$, but a good description of the observational data for the Hubble function up to $z\approx 4$ can be obtained for $\bar{\alpha}=-0.13$.

Now, we can obtain the deviations (in \%) of the dimensionless Hubble parameter $h$ from the $\Lambda$CDM values. At $z=1$ for $\bar{\alpha}=-0.13$ and  $\bar{\alpha}=-0.45$, we have $-1.34\%$, and $-1.91\%$, respectively. However, at $z=3$ the deviations become more significant, and can be obtained as $-4.57\%$ and $-6.54\%$ respectively.

A comparable situation can be seen in the description of the deceleration parameter, shown in the right panel of Fig.~\ref{hq1p}, whose evolution is strongly dependent on the model parameters even at low redshifts. However, $\bar{\alpha}=-0.13$ gives a very good description of the $\Lambda$CDM model values, with the differences slightly increasing in the low redshift region. Hence, for $\bar{\alpha}=-0.13$,  the present model gives a good description of the observational data for the expansion rate, and of the deceleration parameter up to a redshift of $z\approx 4$.

\subsection{The additive case: $f(R,L_m,T)=R+\alpha k(R)+\beta g(T)+2\gamma h(L_m)+\lambda$ }

Let us consider now the case when the Lagrange density of the gravitational field is given by an additive type algebraic structure of the form
\be
f\left(R,L_m,T\right)=R+\alpha\, k(R)+\beta \,g(T)+2\gamma\, h\left(L_m\right)+\lambda,
\ee
where $k(R)$, $g(T)$ and $h\left(L_m\right)$ are arbitrary functions of their arguments.
With this choice of $f$ the cosmological evolution equations can be written as
\begin{align}
-2&(1+\alpha k^\prime)\dot{H}\nonumber\\&=(8\pi+\beta g^\p+\gamma h^\p)(\rho+p)+\alpha\, a\f{d}{dt}\left(\f{1}{a}\dot{k}^\p\right),
\end{align}
\begin{align}
-3&(1-\alpha k^\p)(\dot{H}+2H^2)\nonumber\\&=\alpha k+\beta g+2\gamma h+\lambda-2L_m(\beta g^\p+\gamma h^\p)\nonumber\\&-\f12(\rho-3p)(8\pi+\beta g^\p+\gamma h^\p)+\f{3\alpha}{2a^3}\f{d}{dt}\left(a^3\dot{k}^\p\right),
\end{align}
and
\begin{align}
(8&\pi+\beta g^\p+\gamma h^\p)(\dot{\rho}+\dot{p}+4H(\rho+p))\nonumber\\&=2(\dot{H}+2H^2)(1+\alpha k^\p)\f{d}{dt}\ln\f{1+\alpha k^\p}{6(\dot{H}+2H^2)}\nonumber\\&-(\rho+p)(\beta\dot{g}^\p+\gamma\dot{h}^\p)-\alpha\f{d}{dt}\left(\f{1}{a^3}\f{d}{dt}\left(a^3\dot{k}^\p\right)\right),
\end{align}
respectively.

In the following we will adopt for the additive gravitational Lagrange density the simple form corresponding to $k(R)=0$, $g(T)=T$, and $h\left(L_m\right)=L_m$, so that
\be
f\left(R,L_m,T\right)=R+\beta\, T+2\gamma\, L_m.
\ee

 As for the choice of the matter Lagrangian, we will restrict our analysis to the case $L_m=-\rho$ only.

\subsubsection{The radiation dominated Universe}

By assuming that the matter content of the Universe is radiation with $p=\rho/3$, the generalized Friedmann equations take the form
\be\label{124}
3 \dot{H}+2   (\beta +\gamma +8 \pi )\rho=0,
\ee
\be\label{124a}
3 \dot{H}+6 H^2+\lambda+2 \beta  \rho =0,
\ee
while the evolution of the energy density can be obtained from the energy balance equation as
\begin{align}
(\gamma +8 \pi ) \dot{\rho}+4  (\beta +\gamma +8 \pi )H \rho =0.
\end{align}

One should note that the energy balance equation is not independent of the field equations \eqref{124} and (\ref{124a}).
From Eqs.~ \eqref{124} and (\ref{124a}) one can obtain the time evolution of the Hubble function and of the energy density as follows
\begin{align}\label{127}
H=\sqrt{\f{\lambda}{6} } \tan(g_1(t)),
\end{align}
and
\begin{align}
\rho=\frac{\lambda }{2 (\gamma +8 \pi )} \sec ^2(g_1(t)),
\end{align}
where
\be
g_1(t)=\sqrt{\lambda }
\left(\f{c_1}{2}-\frac{ \sqrt{2}  (\beta +\gamma +8 \pi
	)t}{\sqrt{3}(\gamma +8 \pi) }\right),
\ee
and $c_1$ is a constant of integration. By using Eq. \eqref{127} one can obtain the scale factor as
\begin{align}
a(t)=	c_2 \left[\cos(g_1(t))\right] ^{\frac{\gamma +8 \pi }{2 (\beta +\gamma +8 \pi )}},
\end{align}
where $c_2$ is a constant of integration.

\subsubsection{The dust Universe}

For the additive choice of the Lagrangian of the $f(R,Lm,T)$ theory, with $L_m=-\rho$, and for the equation of state  $p=0$, the metric equations and the energy balance equations have the from
\begin{align}
&2 \dot{H}+ (\beta +\gamma +8 \pi )\rho =0,\\&
6 \dot{H}+12 H^2+2 \lambda+(\beta   -\gamma   -8 \pi)  \rho=0,\\&
(\beta +2 \gamma +16 \pi ) \dot{\rho}+6  (\beta +\gamma +8 \pi )H \rho =0.
\end{align}

These equations have the following exact solutions for the Hubble parameter and matter energy density,
\begin{align}
H=\sqrt{\f{\lambda }{6}} \tan(g_2(t)) ,
\end{align}
and
\begin{align}
\rho=\frac{\lambda }{\beta +2 \gamma +16 \pi } \sec ^2(g_2(t)),
\end{align}
respectively, where
\be
g_2(t)= \sqrt{\lambda }
\left(\f{c_1}{2}-\frac{\sqrt{6}  (\beta +\gamma +8 \pi )\,t}{2\beta +4
	(\gamma +8 \pi )}\right),
\ee
and $c_1$ is an integration constant. Now one can obtain the scale factor as
\begin{align}
	a=c_2 \left[\cos(g_2(t))\right] ^{\frac{\beta +2 (\gamma +8 \pi )}{3 (\beta +\gamma +8
			\pi )}},
\end{align}
where $c_2$ is an integration constant.

\subsubsection{Numerical solutions}

In this Section we consider the numerical solutions for the additive choice of the Lagrangian in the presence of dust and radiation, as the matter constituents of the Universe. We rescale the parameters that enter in the gravitational Lagrangian and the cosmological equations according to
\be
\bar{\beta }=\kappa^2\beta,\quad \bar{\gamma }=\kappa^2\gamma,\quad \lambda=3H_0 \bar{\lambda}.
\ee
 In terms of the redshift the generalized cosmological equations are
\begin{align}\label{ad11}
2 (1+z)h\, h'-  &\left(6 \bar{\gamma }+6 \bar{\beta }+3\right)\Omega _m\nonumber\\&-\left(8
\bar{\gamma }+8 \bar{\beta }+4\right) \Omega _r =0,
\end{align}
and
\begin{align}\label{133}
2 (1+z)h\, h'-&4 h^2-2\bar{\lambda}-8 \bar{\beta } \Omega _r\nonumber\\&+\left(2 \bar{\gamma }-2
\bar{\beta }+1\right)\Omega _m =0,
\end{align}
respectively, while the conservation equations for dust and radiation are
\begin{align}
&(1+z) \left(1+2 \bar{\gamma }\right) \Omega _r'-4 \Omega _r \left(1+2 \bar{\gamma }+2
\bar{\beta }\right)=0,\nonumber\\&
(1+z) \left(1+2 \bar{\gamma }+\bar{\beta }\right) \Omega _m'-3 \Omega _m
\left(1+2 \bar{\gamma }+2 \bar{\beta }\right)=0.
\end{align}

One should note that the value of the parameter $\bar{\lambda }$ can be obtained by the use of Eqs. \eqref{ad11} and \eqref{133} at $z=0$ in terms of $\bar{\beta }$ and $\bar{\gamma}$ as
\begin{align}
\bar{\lambda}=-2+2\left(1+\bar{\beta}+2\bar{\gamma}\right)\Omega_{m0}+2\left(1+2\bar{\gamma}\right)\Omega_{r0}.
\end{align}

Also the current value of the deceleration parameter in this case is
\begin{align}
q_0=-1+\f12\left(1+2\bar{\beta}+2\bar{\gamma}\right)\left(3\Omega_{m0}+4\Omega_{r0}\right).
\end{align}

\begin{figure*}[htbp]
	\includegraphics[scale=0.95]{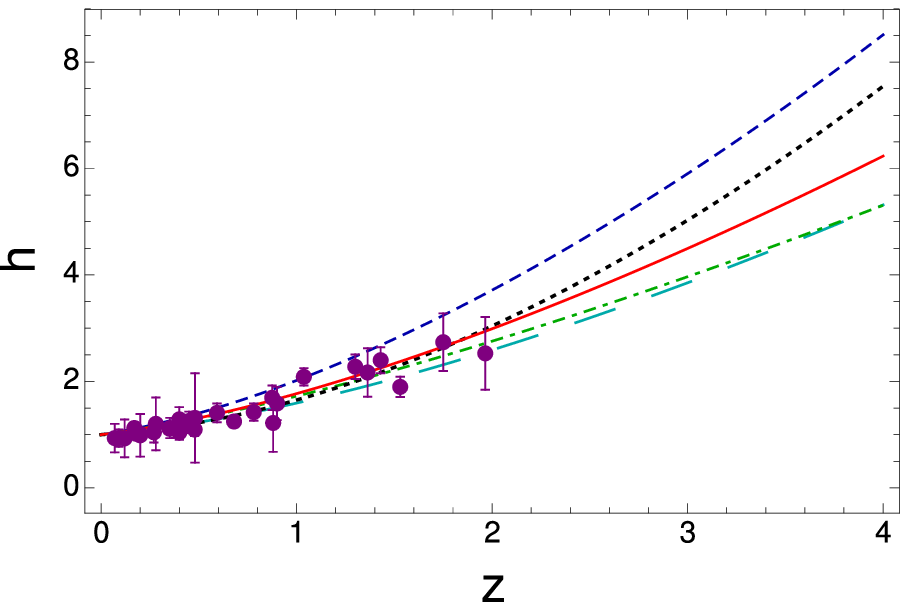}\hspace{.4cm}
	\includegraphics[scale=0.95]{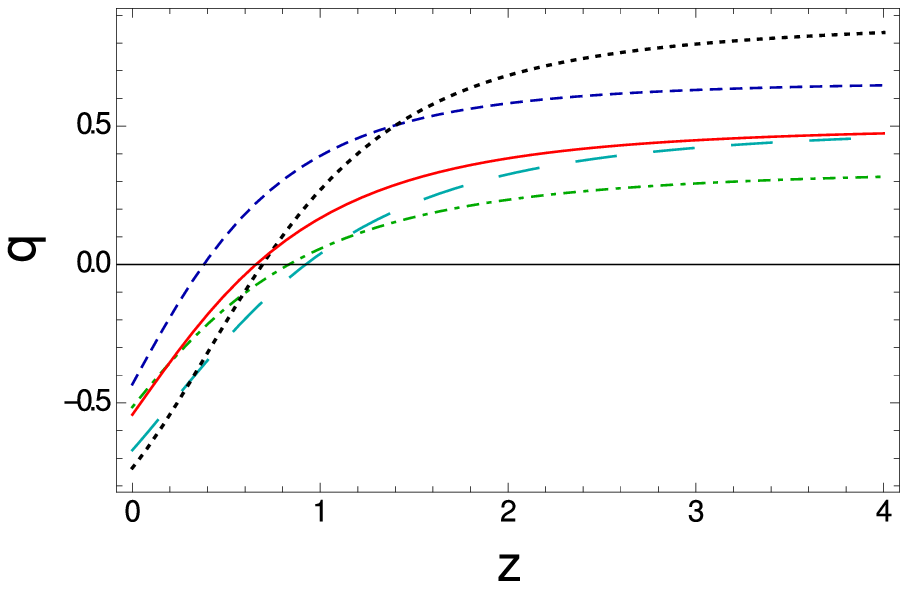}
	\caption{The dimensionless Hubble function (left panel) and the deceleration parameter (right panel) as a function of the redshift $z$ for the additive case of the $f\left(R,L_m,T\right)$ gravity theory, with $L_m=-\rho$, for four different values of the constants $\bar{\beta }$ and $\bar{\gamma}$: $\bar{\beta }=0.12$ and $\bar{\gamma }=0.0$ (dashed curve), $\bar{\beta }=0.0$ and $\bar{\gamma }=-0.14$ (long-dashed curve), $\bar{\beta }=-0.12$ and $\bar{\gamma }=0.15$ (dot-dashed curve), and $\bar{\beta }=0.11$ and $\bar{\gamma }=-0.32$ (dotted curve), respectively. For the Hubble parameter, we have also plotted the observational data together with their errors \cite{hubble}. The $\Lambda$CDM curve is depicted as a red solid curve.\label{hq2ro}}
\end{figure*}

The results obtained by the numerical integration of the generalized Friedmann equations for this case are shown, for four different choices of $\bar{\beta }$ and $\bar{\gamma}$,  in Figs.~\ref{hq2ro}.

As one can see from the left panel of Fig.~\ref{hq2ro}, for some model parameter values the simple additive algebraic structure of the $f\left(R,L_m,T\right)$ gravity theory  provides a good description of the behavior of the Hubble function in the redshift range $z<1$. For higher redshifts a strong dependence on the model parameters does appear, and the differences with respect to the standard $\Lambda$CDM model become important.

For the case with  $\bar{\beta }=-0.12$ and $\bar{\gamma}=0.15$, the percent  deviation of the dimensionless Hubble parameter $h$ with respect to the $\Lambda$CDM values at $z=1$ and $z=3$ are $2.66\%$, and $11.80\%$, respectively. One can also compare these values  with the corresponding quantities for the parameters $\bar{\beta }=0.11$ and $\bar{\gamma}=-0.32$, which are $6.80\%$, and $-11.80\%$, respectively.

An analogues strong dependence on the numerical values of $\bar{\beta }$ and $\bar{\gamma }$ can be seen in the behavior of the deceleration parameter, as shown in the left panel of Fig.~\ref{hq2ro}, with some combinations of the parameters providing a good description of the $\Lambda$CDM data at low redshifts, and with other combinations describing better the standard cosmological model at higher redshifts.

\section{Discussions and final remarks}\label{7}

In the present paper we have considered a generalized gravity model with the
gravitational Lagrangian density given by an arbitrary function of the Ricci
scalar, of the trace of the energy-momentum tensor $T$, and of the Lagrange
density of the matter $L_m$, respectively. The proposed action represents a non-trivial
extension of the standard Hilbert-Einstein action for the standard general relativistic gravitational field, $%
S=\int \left[ R/2+L_{m}\right] \sqrt{-g}\,d^{4}x$, and it unifies two already well-developed modified gravity theories, the $f\left(R,L_m\right)$ theory, and the $f(R,T)$ theory, respectively. It is important to note that the classical
Hilbert-Einstein action, as well as most of its generalizations, has an
\textit{intrinsic additive} algebraic structure, and is constructed as the \textit{sum} of independent
geometrical and physical terms, without involving any coupling between matter and geometry.

On the other hand, the $f\left(R,L_m,T\right)$ modified theory of gravity considered in the present paper
offers the possibility of going beyond this simple algebraic structure, and also brings new degrees of freedom into the gravitational theories. Moreover, besides its theoretical interest, such a generalization may cure some of the cosmological pathologies of either $f(R,T)$ or $f\left(R,L_m\right)$ theories, like, for example, those pointed out in \cite{fT12a}, by providing a description of the cosmological evolution valid from high to low redshifts. In our study we have restricted our numerical analysis of the cosmological evolution for a redshift range $0\leq z\leq 3$, and at higher redshifts some differences between the considered models do appear. To clarify the nature of these deviations it is necessary to perform a fit of the model predictions and observational data. But our preliminary results already indicate that the theory can provide an acceptable description to the observational data, also representing an attractive alternative to the $\Lambda$CDM cosmology.

When formulating the $f\left(R,L_m,T\right)$ gravity theory an interesting analogy with the geometry-matter symmetric Einstein theory is found. It is known that the field equations (\ref{eqsym}) contain the cosmological constant as a constant of integration \cite{La1,La2,La3,La3a,La3b}. This can be shown immediately by taking the divergence of Eqs.~(\ref{eqsym}), together with the postulates of the conservation of the matter energy-momentum tensor, $\nabla _{\mu}T^{\mu \nu}=0$, which gives $\nabla _{\mu}R=-8\pi \nabla_{\mu}T$, $\nabla _{\mu}\left(R+8\pi \right)=0$, and $R+8\pi T=\Lambda$, where $\Lambda $ is an integration constant. After substitution of $R=\Lambda-8\pi T$ into Eq.~(\ref{eqsym}) we reobtain the standard Einstein equations in the presence of a cosmological constant, which thus becomes a simple integration constant. This representation of the Einstein field equations is also called unimodular gravity. The trace-free Einstein equations can be obtained from the variational principle \cite{HeTe,La4}
\be
S\left[g,W,\Lambda\right]=\frac{1}{\kappa ^2}\int{\left[-\frac{R}{2}+\Lambda\left(\nabla _{\mu}W^{\mu}-1\right)\right]\sqrt{-g}d^4x} .
\ee

The field equations of the traceless representation of the $f\left(R,L_m,T\right)$ gravity theory represent an extension of the geometry-matter symmetric Einstein theory, with the field equations taking the form
\be\label{eq138}
R_{\mu \nu}-\frac{1}{4}Rg_{\mu \nu}=8\pi \tilde{G}_{eff}\left(T_{\mu \nu}-\frac{1}{4}Tg_{\mu \nu}\right)+U_{\mu \nu},
\ee
where
\be
\tilde{G}_{eff}=\frac{1}{f_R}\left(1+\frac{f_L+2f_T}{16\pi}\right),
\ee
and
\be
U_{\mu \nu}=- \frac{1}{f_{R}}\left( \frac{1}{4}g_{\mu \nu }\Box -\nabla _{\mu }\nabla
_{\nu }\right) f_{R} +\frac{f_T}{f_R}\left(\tau _{\mu \nu}-\frac{1}{4}\tau g_{\mu \nu}\right),
\ee
respectively. In the absence of the electromagnetic type interactions, the supplementary term $U_{\mu \nu}$ generated by the geometry-matter coupling takes the simple form $U_{\mu \nu}=- \left(1/f_{R}\right)\left[ \left(g_{\mu \nu }/4\right)\Box -\nabla _{\mu }\nabla
_{\nu }\right] f_{R} $.

As one can see from the field equations (\ref{eq138}), the matter energy-momentum tensor $T_{\mu \nu}$ is generally not conserved, and $\nabla _{\mu}T^{\mu \nu}\neq 0$. Of course one can construct conservative models in which the energy-momentum conservation is imposed a priori, $\nabla _{\mu}T^{\mu \nu}=0$, but the possible non-conservation of $T_{\mu \nu}$ has far reaching physical implications. First of all, the motion of free particles is not geodesic anymore, and it takes place in the presence of an extra-force generated by the geometry-matter coupling. The extra-force generates an extra-acceleration, and hence, in the Newtonian limit, the total acceleration $\vec{a}$ of a massive gravitating objects can be written as $\vec{a}=\vec{a}_N+\vec{a}_E$. If $\vec{a}_E=0$, then the acceleration of the object is $\vec{a}=\vec{a}_N$, with the Newtonian acceleration of an object moving in the gravitational field created by a mass $M$ given by $\vec{a}_N=-GM\vec{r}/r^3$. From the acceleration equation we obtain
\be
\vec{a}_E\cdot \vec{a}_N=\frac{1}{2}\left(\vec{a}^2-\vec{a}_N^2-\vec{a}_E^2\right),
\ee
which allows to express the vector $\vec{a}_N$ as
\be\label{eq142}
\vec{a}_N=\frac{1}{2}\left(\vec{a}^2-\vec{a}_N^2-\vec{a}_E^2\right)\frac{\vec{a}}{\vec{a}_E\cdot \vec{a}}+\vec{C}\times \vec{a}_E,
\ee
where $\vec{C}$ is an arbitrary vector, which for simplicity will be taken as zero in the following.  From Eq.~(\ref{eq142}) it follows that in order to obtain a consistent mathematical description  $\vec{a}_E\cdot \vec{a}\neq 0$, that is, the vectors $\vec{a}_E$ and $\vec{a}$ cannot be orthogonal. Without any loss of generality we will assume that they are parallel. By denoting
\be
\frac{1}{\tilde{a}_E}=\frac{1}{2}\frac{\left|\vec{a}\right|}{\left|\vec{a}_E\right|}\left(1-\frac{\vec{a}_N^2}{\vec{a}^2}-\frac{\vec{a}_E^2}{\vec{a}^2}\right),
\ee
we obtain
\be\label{eq144}
\vec{a}=\tilde{a}_E\vec{a}_N,
\ee
that is, the total acceleration of an object in the gravitational field of a mass $M$ is proportional to the Newtonian acceleration. Such a relation has been already inferred from the study of the galactic rotation curves of massive test particles gravitating around the galactic center, and it is called the radial acceleration relation (RAR) \cite{Rar1,Rar2,Rar3, Rar4}. The radial acceleration relation is an empirical relation that tries to establish a connection between the centripetal acceleration $a_{obs} (R)= V_ {rot}^2 (R)/R$ observed in galaxies, and usually attributed to the presence of dark matter, and the Newtonian acceleration $a_{bar}(R)=V_{bar}^2/R$ of the observed baryonic matter distribution. The relation can be formulated as
\be
a_{obs}=\nu \left(\frac{a_{bar}}{a_+}\right)a_{bar},
\ee
where $\nu(x)$ denotes a fitting function whose mathematical form must be determined observationally, while $a_+$ denotes an acceleration scale. A similar relation, indicating the proportionality of the total acceleration to the Newtonian one does naturally appear in the low velocity/low density limit of the $f\left(R,L_m,T\right)$ gravity, as shown by Eq.~(\ref{eq144}), where such a relation is a direct consequence of the geometry-matter coupling. More interestingly, the extra-acceleration predicted by the theory in the Newtonian limit is dependent on the density of the baryonic matter, $a_E=a_E(\rho)$, a result that indicates a possible relation with the chameleon field models \cite{Cam1,Cam2, Cam3}. The chameleon scalar field is a light scalar field, with a mass depending on the background baryonic matter density. However, such environmental effects can also be generated by the geometry-matter coupling, indicating a baryonic density dependence of the astrophysical processes on large scales.

We have also obtained the generalized Poisson equation in the Newtonian limit of the theory, by using a series expansion near a fixed point of the theory, and restricting our analysis to the first order terms. The series expansion also generates a constant term, that can be interpreted as the cosmological constant, and which in the present approach is nothing but the difference of the Lagrangian density at a fixed point at its first order approximation, $\Lambda =\left(1/2\alpha \right)\left[f\left(R_0,T_0,L_{m0}\right)- \alpha R_0-\beta T_0-\gamma L_{m0}\right]$. Hence in this approximation the cosmological constant appears as a second order effect in the series expansion of the full Lagrangian density $f\left(R,L_m,T\right)$, and hence this interpretation may justify its smallness. In the same order of approximation there is a shift of the gravitational constant, determined by the derivatives of $f$ with respect to $L_m$ and $T$ estimated for some background values of the model parameters. However, the constraints on the numerical value of the gravitational constant \cite{Gr1,Gr2} could lead to the determination of some limits of the gravitational Lagrangian in modified gravity with geometry-matter coupling.

The Dolgov-Kawasaki instability is a general feature of modified theories of gravity, in which higher order equations of motion do emerge that may exhibit unphysical behaviors. For example, unitarity may be broken, and ghosts or tachyons could develop. Moreover, the breaking of the laws of gravity in the weak field regime and for small curvatures leads to contradiction with basic observational facts. We have investigated in detail the Dolgov-Kawasaki instability for the $f\left(R,L_m,T\right)$ theory, and we have obtained the criterion for the stability of the theory, $f_{RR}\left(0\right)\geq 0$. This is the standard stability condition as obtained in, for example, $f(R)$ gravity, a condition that assures the instability of the perturbations, and the absence of the tachyonic instabilities.

An important field of application of $f\left(R,L_m,T\right)$ is cosmology. We have obtained the generalized Friedmann equations (\ref{traceless}) and (\ref{trace}), which must be usually considered together with the energy-momentum balance equation (\ref{balance}). The condition for the accelerated expansion is given by Eq.~(\ref{eq87}), and it can be satisfied by a large variety of functional forms of the Lagrange density of the gravitational field in the $f\left(R,L_m,T\right)$ gravity. We have investigated in detail two distinct classes of cosmological models, corresponding to a multiplicative and an additive structure of the terms involving the geometry-matter coupling. The evolutionary cosmological models have a large variety of behaviors, and different types of accelerating or decelerating models can be constructed. Overall, for some values of the model parameters, the considered model give a good description of the cosmological observational data, and they can reproduce the predictions of the $\Lambda$CDM paradigm at both low ($z<1$) and higher ($z\approx 4$) redshifts.

It should be noted here that in the  plots of the Hubble functions $H(z)$,  we have normalized them to the current value of the Hubble function  $H_0$.  However, for every choice of the model parameters, the current value of the Hubble function is different, and this value should be obtained by fitting the theory with the observational data, thus obtaining some observational constraints on the model parameters. This will be done in future works.  As for the Hubble tension problem, we would like to point out that all modified gravity theories that make the Hubble tension better, will make the $\sigma_8^0$ tension worsen \cite{H0ten1, H0ten2,H0ten3}. However, as a possible solution to these problems, in \cite{decayDM} it was shown that a decaying dark matter model can be considered as a solution to both tensions simultaneously. Such an approach can also be extended to gravitational theories with geometry-matter coupling.

The present gravitational theory still faces the problem of the degeneracy of the matter Lagrangian, leading to two different classes of models corresponding to the choices $L_m=-\rho$, and $L_m=p$, respectively.

However, a possible solution of this problem can be obtained as follows. For simplicity we assume that the Lagrangian density $L_{m}$ of the ordinary matter is a function of the metric tensor components $g_{\mu \nu }$ only, and not of its derivatives. Then for the energy-momentum tensor we obtain
the expression $T_{\mu \nu }=L_{m}g_{\mu \nu }-2\partial L_{m}/\partial g^{\mu
\nu }$. Moreover, we assume that the matter satisfies a barotropic equation of state $p=p(\rho)$, and thus the matter Lagrangian, that generally can be considered a function of both $\rho$ and $p$, $L_m=L_m(\rho, p)$ becomes a function of the matter density only, $L_m=L_m(\rho)$. By assuming that the matter current is conserved, $\nabla _{\nu }\left( \rho u^{\nu }\right) =0$, and by taking into account the relation $\delta \rho
=-\left( 1/2\right) \rho \left( g_{\mu \nu }+u_{\mu }u_{\nu }\right) \delta
g^{\mu \nu }$ \cite{fLm4}, it follows that the energy-momentum tensor can be written in the form
\begin{equation}
T^{\mu \nu }=\rho \frac{dL_{m}}{d\rho }u^{\mu }u^{\nu }+\left( L_{m}-\rho
\frac{dL_{m}}{d\rho }\right) g^{\mu \nu }.
\end{equation}
With the use of the mathematical identity $u^{\nu }\nabla _{\nu }u^{\mu }=d^{2}x^{\mu }/ds^{2}+\Gamma _{\nu
\lambda }^{\mu }u^{\nu }u^{\lambda }$, from the condition $\nabla ^{\mu }T_{\mu \nu }=0$ we obtain the equation of motion of the fluid as
\begin{equation}\label{151}
\frac{d^{2}x^{\mu }}{ds^{2}}+\Gamma _{\nu \lambda }^{\mu }u^{\nu }u^{\lambda
}+h^{\mu \nu}\nabla _{\nu}\ln \left|K\frac{dL_m(\rho)}{d\rho}\right|=0,
\end{equation}
where $K$ is a constant of integration. A comparison of Eqs.~(\ref{151}) and (\ref{eqmot}) gives the relation
\be\label{152}
K\frac{dL_m(\rho)}{d\rho}=\sqrt{Q}\approx 1+U\left( \phi,\rho ,p,T, L_m,f_m, \nabla _{\nu}T,...\right),
\ee
where we have also used Eq.~(\ref{41}). The above equation leads us to the general result that in modified gravity theories with geometry-matter coupling the matter energy-momentum tensor must be constructed dynamically for each individual model, by solving the differential equation (\ref{152}). In the first order of approximation, when $U\left( \phi,\rho ,p,T, L_m,f_m, \nabla _{\nu}T,...\right)\ll1$, Eq.~(\ref{152}) gives $L_m\approx \rho /K$, which is the (approximate) expression we have adopted in our investigations of the cosmological models, and which coincides with the standard general relativistic matter Lagrangian. This limiting procedure also fixes the value of the constant $K$ as $K=-1$.

To conclude, the current investigation of the unified $f\left(R,T,L_m\right)$ gravity theory opens
new avenues for the study of the theoretical, observational and
even experimental aspects of the large class of alternative theories of gravity in the presence of
geometry-matter coupling. In addition, the basic results obtained in the present work may also contribute to the better understanding of the physical and mathematical structure of other modified gravity theories.

\section*{Acknowledgments}

We would like to thank to the anonymous referee for comments and suggestions that helped us to significantly improve our manuscript. T. H. would like to thank to the Yat Sen School of the Sun Yat Sen University in Guangzhou, P. R. China, for the kind hospitality offered during the early stages of the preparation of this work.

\end{document}